\documentclass{article}

\usepackage[main, final]{neurips_2026}

\usepackage[utf8]{inputenc} % allow utf-8 input
\usepackage[T1]{fontenc}    % use 8-bit T1 fonts
\usepackage{hyperref}       % hyperlinks
\usepackage{url}            % simple URL typesetting
\usepackage{booktabs}       % professional-quality tables
\usepackage{amsfonts}       % blackboard math symbols
\usepackage{nicefrac}       % compact symbols for 1/2, etc.
\usepackage{microtype}      % microtypography
\usepackage{xcolor}         % colors

\usepackage{amsmath}
\usepackage{amssymb}
\usepackage{mathtools}
\usepackage{amsthm}
\usepackage{float}
\usepackage{tikz}
\usetikzlibrary{positioning}
\usepackage{caption}

\newcommand{\ket}[1]{\lvert #1 \rangle}
\newcommand{\bra}[1]{\langle #1 \lvert}

\usepackage{wrapfig}
\usepackage{tabularx}

\usepackage{colortbl}

\usepackage{multirow}

\usepackage[capitalize,noabbrev]{cleveref}

\title{Structured Unitary Tensor Network Representations for Circuit-Efficient Quantum Data Encoding}

\author{%
  Guang Lin \\
  RIKEN AIP\\
  \texttt{guang.lin@riken.jp} \\
  \And
  Toshihisa Tanaka \\
  Tokyo University of Agriculture and Technology \\
  \texttt{tanakat@cc.tuat.ac.jp} \\
  \And
  Qibin Zhao\thanks{Corresponding author} \\
  RIKEN AIP \\
  \texttt{qibin.zhao@riken.jp} \\
}

\begin{document}

\maketitle

\begin{abstract}
Encoding classical data into quantum states is a central bottleneck in quantum machine learning: many widely used encodings are circuit-inefficient, requiring deep circuits and substantial quantum resources, which limits scalability on quantum hardware. In this work, we propose tensor network-based quantum data encoding (TNQE), a circuit-efficient framework built on structured unitary tensor network (TN) representations. TNQE first represents each classical input via a TN decomposition and then compiles the resulting tensor cores into an encoding circuit through two complementary core-to-circuit strategies. To make this compilation trainable while respecting the unitary nature of quantum operations, we introduce a unitary-aware constraint that parameterizes TN cores as learnable block unitaries, enabling them to be directly optimized and directly encoded as quantum operators. The proposed TNQE framework enables explicit control over circuit depth and qubit resources, allowing the construction of shallow, resource-efficient circuits. Across a range of benchmarks, TNQE achieves encoding circuits as shallow as $0.04\times$ the depth of amplitude encoding, while naturally scaling to high-resolution images ($256 \times 256$) and demonstrating practical feasibility on real quantum hardware.
\end{abstract}

\section{Introduction}

In recent years, quantum computing has rapidly advanced from early prototypes with only a few qubits to modern quantum processors with hundreds of qubits \citep{ladd2010quantum,ibmblog2023}. By exploiting quantum mechanical phenomena such as superposition and entanglement, quantum computers enable information representations and computational paradigms that differ fundamentally from classical systems \citep{friedman2000quantum,horodecki2009quantum}. These properties suggest the potential for advantages in specific tasks, including the compact representation of high-dimensional data using relatively few qubits \citep{devadas2025quantum}. Motivated by this potential, quantum machine learning (QML) has emerged as an interdisciplinary field that seeks to integrate quantum computing with machine learning in order to extend classical learning capabilities \citep{rodriguez2025survey,corli2025quantum}.

Despite rapid progress, QML remains at an early stage of development, with many fundamental challenges still unresolved \citep{brunet2024quantum}. Among these challenges, the efficient encoding of classical data into quantum systems stands out as a central and pervasive bottleneck \citep{bermejo2024quantum}. As the entry point from classical information to quantum states, quantum data encoding determines the initial quantum state and the structure of the resulting quantum circuit, thereby shaping all subsequent operations. Consequently, the choice of encoding strategy has a direct impact on key practical factors \citep{munikote2024comparing}. Existing quantum data encoding strategies \citep{weigold2020data,placidi2023mnisq} often rely on complex and deep circuits to represent classical data, leading to substantial resource overhead and error accumulation \citep{bose2025qubit}. These limitations severely constrain the scalability and applicability of quantum machine learning.

\begin{figure}[t]
\begin{center}
\includegraphics[width=\linewidth]{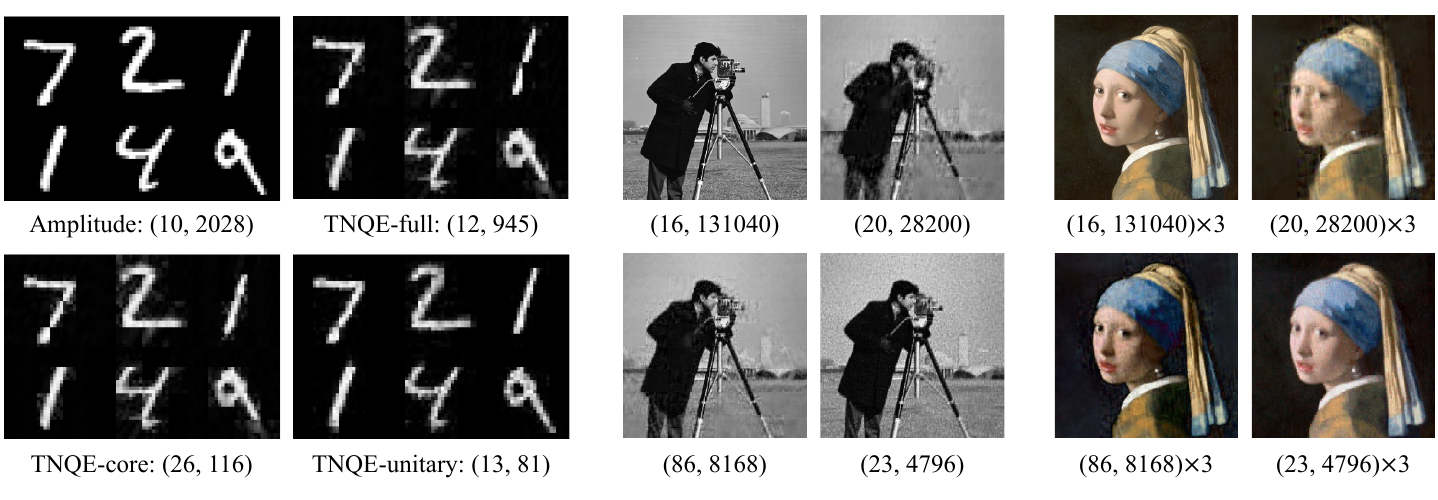}
% \vskip -5pt
\caption{Different-resolution visual comparisons in simulation are conducted using $32 \times 32$, $256 \times 256$, and $256 \times 256 \times 3$ images. The ordering of methods is consistent across resolutions, and the circuit structure is reported as (qubit, depth). Amplitude encoding enables accurate encoding in simulation at the cost of deep circuits, TNQEs provide different trade-offs between qubit and depth.}
\label{fig:vis_comparison}
\end{center}
% \vskip -9pt
\end{figure}

To address these challenges, we propose TNQE, a tensor network (TN)-based framework for circuit-efficient quantum data encoding that constructs executable quantum circuits from structured tensor representations of classical data.
TNs provide compact factorizations of high-dimensional arrays into collections of low-rank tensor cores \citep{kolda2009tensor,cichocki2015tensor}, offering an explicit handle to trade off representation fidelity and model complexity.
Crucially, TNs are also deeply connected to quantum systems, as they can describe quantum states and local operations \citep{orus2014practical}.
These properties make TNs a natural intermediate language for encoding classical data to quantum circuits in a structured and efficient manner.

Building on this perspective, TNQE decouples encoding into (i) a TN decomposition of the classical data and (ii) a core-to-circuit conversion step that turns tensor cores into unitaries.
We instantiate two complementary conversion strategies, including TNQE-full that converts right-canonical MPS cores into isometries and then completes them into local unitaries, yielding a sequential circuit realization; and TNQE-core that prepares each tensor core using a dedicated sub-circuit, avoiding inter-core entangling operations and enabling a shallow, modular, and parallelizable encoding.
In addition, we introduce TNQE-unitary, which incorporates a unitary-aware optimization constraint: tensor cores are parameterized directly as learnable block unitaries, so they can be directly optimized and directly encoded as quantum operators without post hoc unitary conversion or synthesis.

We benchmark TNQE across circuit depth, qubit count, operation count, and encoding fidelity.
Across the settings, TNQE preserves informative structure with substantially shallower circuits, achieving depths as low as $0.04\times$ that of amplitude encoding.
Notably, while many existing methods for high-resolution images remain largely theoretical \citep{lisnichenko2023quantum,brunet2024quantum}, TNQE successfully encodes $256\times256$ images into quantum circuits in simulation (\cref{fig:vis_comparison}).
We further report initial experiments on IBM quantum processors, where the shallow TNQE circuits retain image structure under hardware noise while deep exact encodings do not.
Together, these results indicate that TNQE offers a scalable and practical framework for circuit-efficient quantum data encoding.
Our main contributions are as follows:
\begin{itemize}
    \item We propose TNQE, a TN-based quantum data encoding that bridges classical data and executable quantum circuits via structured core-to-circuit conversion.
    \item We develop three instantiations: TNQE-full and TNQE-core as sequential and parallel realizations, and TNQE-unitary, our main contribution, which imposes unitary-aware constraints to parameterize the cores as learnable blocks for direct optimization and execution.
    \item We demonstrate the scalability and practicality of TNQE through extensive experiments, showing substantial depth reductions (down to $0.04\times$ vs.\ amplitude encoding), successful encoding of high-resolution images, and feasibility on real quantum hardware.
\end{itemize}
% \footnotetext{https://quantum.cloud.ibm.com}

\section{Related works}
\noindent\textbf{Tensor networks}
provide an efficient framework for modeling and processing high-dimensional data through compact structured representations \citep{kolda2009tensor,cichocki2015tensor}. Common tensor network formats include matrix product state \citep[MPS,][]{dilip2022data} and tensor train \citep[TT,][]{oseledets2011tensor}, as well as specific structures inspired by quantum, such as quantized tensor train \citep[QTT,][]{khoromskij2011d}. Notably, these TNs have been widely used in various applications, including both quantum systems \citep{orus2019tensor,berezutskii2025tensor} and classical machine learning \citep{sengupta2022tensor,shin2024dequantizing}. Therefore, TNs provide a potential bridge between classical data and quantum systems, inspiring our TNQE framework for quantum data encoding.

\noindent\textbf{Quantum data encoding} is a fundamental step in quantum machine learning that encodes classical data into quantum states \citep{schuld2014quest}.
Early quantum data encoding strategies, including basis encoding and amplitude encoding \citep{schuld2019quantum}, originate from quantum information processing and quantum algorithm design. However, these encodings often lead to complex circuits that are challenging to implement for image data. Recent works have proposed more practical encoding methods for QML. \citet{dilip2022data,Jobst2024} construct sequential circuits using 1-D MPS  and reduce complexity by partitioning images into $p-$patches. However, this strategy disrupts spatial structure and rapidly increases complexity as resolution grows.
\citet{ranga2024hybrid,lu2025fidelity} use convolutional neural networks (CNNs) to extract low-dimensional feature vectors in the classical system and then encode these vectors into quantum circuits. While supporting arbitrary image resolutions, these quantum-classical hybrid methods operate on extracted features instead of full information (FI), limiting quantum machine learning as discussed in \Cref{app:discussion}.

\begin{wraptable}{r}{0.51\linewidth}
\vskip -11pt
  \centering
\caption{Comparison of quantum data encoding methods in terms of scalability and resource requirements. Here $s^2$ denotes the $s \times s$ image size, and $r$ and $N_\ell$ are hyperparameters of TNQE.}
% \vskip 9pt
\resizebox{\linewidth}{!}{
\begin{tabular}{l@{\hspace{1pt}}c@{\hspace{8pt}}c@{\hspace{3pt}}c@{\hspace{5pt}}c}
\toprule
Methods & Qubits & Depth & Image size & FI \\
\midrule
Basis encoding & $O(s^2)$ & $O(1)$ & \textcolor[rgb]{1,0,0}{limited} & \textcolor[rgb]{0,0.6,0}{\checkmark} \\
Amplitude encoding & $O(\log s^2)$ & $O(s^2)$ & \textcolor[rgb]{1,0,0}{limited} & \textcolor[rgb]{0,0.6,0}{\checkmark} \\
MPS-based encoding & $O(p (\log s^2/p))$ & $O(s^2/p)$ & \textcolor[rgb]{1,0,0}{limited} & \textcolor[rgb]{0,0.6,0}{\checkmark} \\
CNN-based hybrid & $O(d)$ & $O(1)$ & \textcolor[rgb]{0,0.6,0}{arbitrary} & \textcolor[rgb]{1,0,0}{$\times$} \\
TNQE-full & $O(\log s^2 + \log r)$ & $O(4 r^2 \log s)$ & \textcolor[rgb]{1,0,0}{limited} & \textcolor[rgb]{0,0.6,0}{\checkmark} \\
TNQE-core & $O(\log s \log r)$ & $O(4 r^2)$ & \textcolor[rgb]{0,0.6,0}{scalable} & \textcolor[rgb]{0,0.6,0}{\checkmark} \\
TNQE-unitary  & $O(\log s^2 + \log r)$ & $O(N_\ell \log s)$ & \textcolor[rgb]{0,0.6,0}{scalable} & \textcolor[rgb]{0,0.6,0}{\checkmark} \\
\bottomrule
\end{tabular}}
\label{tab:comparison_encoding}
\vskip -9pt
\end{wraptable}

As summarized in \cref{tab:comparison_encoding}, conventional methods are constrained by quantum resources and are therefore limited to low-resolution images.
While hybrid methods can handle higher-resolution images, the quantum output must still be processed by a CNN to reconstruct the original image for applications requiring FI quantum representations, undermining the quantum advantages.
In contrast, by leveraging TN representations, our proposed TNQE offers flexible control over depth and qubit requirements, enabling scalability to higher-resolution images.
These properties make TNQE a practical and promising solution for quantum data encoding.

\section{Preliminary}
In this work, we adopt quantized tensor train \citep[QTT,][]{khoromskij2011d} as the underlying tensor network structure. QTT is particularly well suited for image data and quantum data encoding, as it exploits spatial structure through mode quantization \citep{loeschcke2024coarse} and produces tensor cores with physical dimensions, which naturally align with qubit-based quantum circuits.

QTT builds upon the idea of mode quantization, where each physical dimension is recursively decomposed into binary factors, typically powers of two. Specifically, consider a two-dimensional image defined on an $s \times s$ size, where $s = 2^{L}$. Through mode quantization, the spatial dimensions are factorized as $(2_1 \times 2_2 \times \cdots \times 2_L) \times (2_1 \times 2_2 \times \cdots \times 2_L)$:
\tikz[baseline=-0.5ex]{
    \node[draw, circle, inner sep=1pt] (tensor1) {\scriptsize $\mathcal{A}$};
    \node[draw, circle, inner sep=1pt, right=0.3cm of tensor1] (tensor2) {\scriptsize $\mathcal{A}$};
    \node[inner sep=0pt, right=0.3cm of tensor2] (tensor3) {{ }\dots{ }};
    \node[draw, circle, inner sep=1pt, right=0.3cm of tensor3] (tensor4) {\scriptsize $\mathcal{A}$};
    \node[draw, circle, inner sep=1pt, right=0.3cm of tensor4] (tensor9) {\scriptsize $\mathcal{A}$};
    \node[draw, circle, inner sep=1pt, right=0.3cm of tensor9] (tensor10) {\scriptsize $\mathcal{A}$};
    \node[inner sep=0pt, right=0.3cm of tensor10] (tensor11) {{ }\dots{ }};
    \node[draw, circle, inner sep=1pt, right=0.3cm of tensor11] (tensor12) {\scriptsize $\mathcal{A}$};

    \draw (tensor1) -- (tensor2);
    \draw (tensor2) -- (tensor3);
    \draw (tensor3) -- (tensor4);
    \draw (tensor4) -- (tensor9);
    \draw (tensor9) -- (tensor10);
    \draw (tensor10) -- (tensor11);
    \draw (tensor11) -- (tensor12);

    \draw (tensor1) -- ++(0.3,0.3);
    \draw (tensor2) -- ++(0.3,0.3);
    \draw (tensor4) -- ++(0.3,0.3);
    \draw (tensor9) -- ++(-0.3,0.3);
    \draw (tensor10) -- ++(-0.3,0.3);
    \draw (tensor12) -- ++(-0.3,0.3);
}, which corresponds to a $2L$-dimensional hypercube representation.
Following the QTT construction, the binary factors are reordered and grouped according to their hierarchical levels. In particular, the factors corresponding to the same level across different spatial dimensions are grouped together, yielding an ordering of the form
$(2_1^{(x)} \times 2_1^{(y)}) \cdot (2_2^{(x)} \times 2_2^{(y)}) \cdots (2_L^{(x)} \times 2_L^{(y)})$:
\tikz[baseline=-0.5ex]{
    \node[draw, circle, inner sep=1pt] (tensor1) {\scriptsize $\mathcal{A}$};
    \node[draw, circle, inner sep=1pt, right=0.3cm of tensor1] (tensor2) {\scriptsize $\mathcal{A}$};
    \node[inner sep=0pt, right=0.3cm of tensor2] (tensor3) {{ }\dots{ }};
    \node[draw, circle, inner sep=1pt, right=0.3cm of tensor3] (tensor4) {\scriptsize $\mathcal{A}$};

    \draw (tensor1) -- (tensor2);
    \draw (tensor2) -- (tensor3);
    \draw (tensor3) -- (tensor4);

    \draw (tensor1) -- ++(0.3,0.3);
    \draw (tensor1) -- ++(-0.3,0.3);
    \draw (tensor2) -- ++(0.3,0.3);
    \draw (tensor2) -- ++(-0.3,0.3);
    \draw (tensor4) -- ++(0.3,0.3);
    \draw (tensor4) -- ++(-0.3,0.3);
},
where the superscripts indicate the two spatial dimensions ($x,y$) of the image. From the perspective of encoding, the QTT format provides a particularly favorable interface, as each tensor core can be naturally encoded to a local unitary acting on two physical qubits alongside the bond qubits.

\section{Method}
\label{sec:method}

\begin{figure}[t]
\begin{center}
\includegraphics[width=0.98\linewidth]{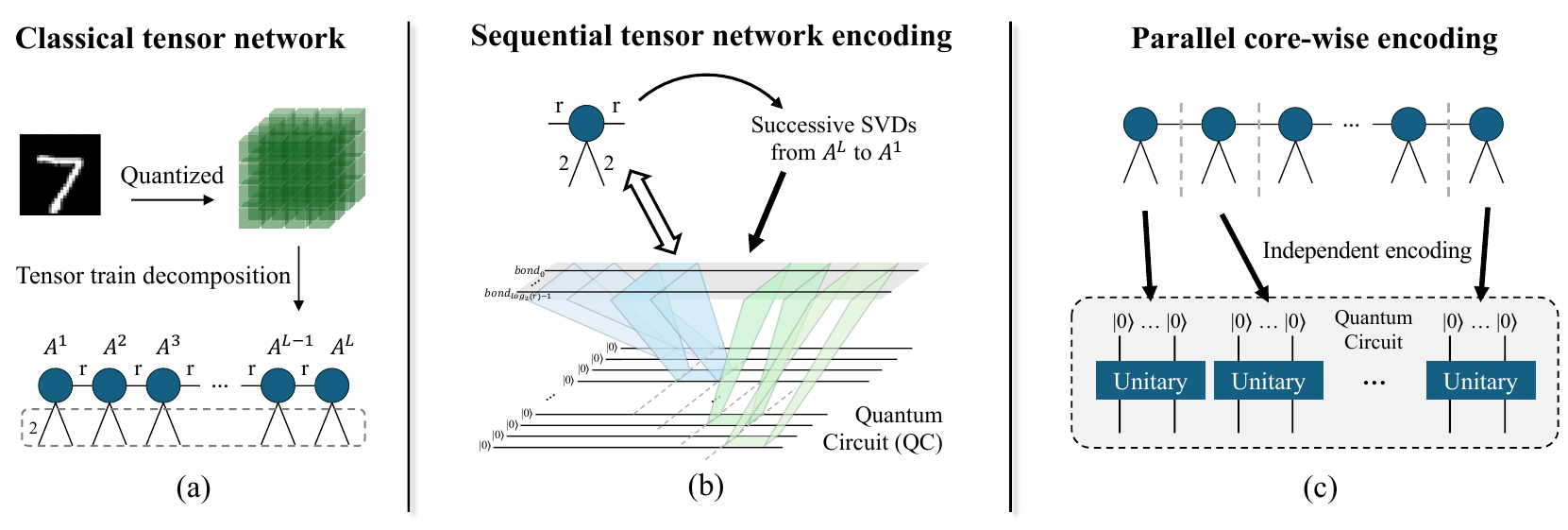}
\caption{Illustration of the proposed TNQEs. First, (a) classical data are represented using QTT. Then, (b) full tensor network encoding and (c) core-wise encoding convert the cores into circuits.}
\label{fig:TNQE}
\end{center}
% \vskip -9pt
\end{figure}

\noindent\textbf{Problem and encoding.}
As illustrated in \cref{fig:TNQE}, our TNQE framework establishes a novel pipeline for transforming classical data into quantum circuits. Specifically, given a compressible image $x\in\mathbb{R}^{s\times s}$ with $s=2^L$, our goal is to construct a shallow executable circuit that prepares the corresponding quantum state. The framework decouples the encoding process into (a) classical tensor network decomposition and (b\&c) two quantum circuit realization pipelines.

The three TNQE variants differ primarily in the role played by the tensor network. TNQE-full and TNQE-core both encode the cores obtained from a conventional QTT decomposition.
TNQE-unitary follows a fundamentally different route: it uses the QTT structure as a prior and learns the tensor cores directly under a unitary-aware constraint. Specifically, it retains the factorization $x \approx \mathcal{A}^{(1)}\mathcal{A}^{(2)}\cdots\mathcal{A}^{(L)}$, but parameterizes each core through a learnable block unitary rather than treating it as an arbitrary tensor.
In the following sections, we describe each in detail.

\subsection{Full tensor network encoding}
\label{subsubsec:full_tn_encoding}

In this strategy, we encode QTT cores to a quantum circuit by viewing the MPS \citep{dilip2022data} as a sequence of isometries that can be completed to unitaries and implemented as quantum operators.

\noindent\textbf{From QTT cores to an MPS.}
First, the QTT representation decomposes the input into a sequence of tensor cores $\mathcal{A}^{(k)} \in \mathbb{C}^{r_{k-1} \times 2 \times 2 \times r_k}, k=1,\dots,L,$ with boundary conditions $r_0=r_L=1$ and $r_k=r$ for $2\le k\le L-1$.
Each core contains two binary physical indices $i_k,j_k\in\{0,1\}$.
We merge them into a single four-level physical index $p_k = 2 i_k + j_k, p_k \in \{0,1,2,3\},$ which converts the QTT core into an MPS tensor with physical dimension $d=4$: $\mathcal{A}^{(k)} \ \mapsto\  A^{(k)} \in \mathbb{C}^{r_{k-1} \times d \times r_k}.$

\noindent\textbf{Right-canonicalization and isometric condition.}
Then, to enable an encoding to quantum operators, we further transform the MPS into a right-canonical form by a sequence of singular value decompositions \citep[SVDs,][]{golub1971singular} applied from right ($A^L$) to left ($A^1$).
Specifically, we reshape each MPS tensor $A^{(k)}$ into a matrix by grouping $(p_k, \alpha_k)$ as $A^{(k)} \ \mapsto\  \mathbf{A}^{(k)} \in \mathbb{C}^{r_{k-1} \times (d\,r_k)}$, perform an SVD, and absorb the non-isometric factor into the neighboring matrix to the left; details are provided in \cref{app:right_canonicalization}. No singular values are discarded in this sweep, so right-canonicalization does not change the represented tensor.
After this procedure, the matrices $\mathbf{A}^{(k)}$ for $k\ge 2$ satisfy the right-isometric  condition $\mathbf{A}^{(k)} (\mathbf{A}^{(k)})^\dagger = I_{r_{k-1}}$.

As a consequence of right-canonicalization, the norm of the global state depends solely on the first matrix $\mathbf{A}^{(1)}$.
Hence, global normalization $\|\psi\|=1$ can be enforced by rescaling $A^{(1)}$ alone, while keeping $\mathbf{A}^{(k)}$ for $k\ge2$ strictly isometric. Each right-canonical matrix can be interpreted as an isometry
$V^{(k)} \equiv (\mathbf{A}^{(k)})^\dagger \in \mathbb{C}^{m \times n}, m = d\,r_k, n = r_{k-1},$
which satisfies $(V^{(k)})^\dagger V^{(k)} = I_n$, as detailed in \cref{app:right_canonicalization}. This isometry maps the incoming bond space to the joint physical and outgoing bond space in a norm-preserving manner.

\noindent\textbf{Local unitaries and quantum circuit construction.}
However, to realize such isometries on a quantum computer, they must be embedded into unitary operations acting on a fixed Hilbert space.
Specifically, given an isometry $V^{(k)}\in\mathbb{C}^{m\times n}$, we construct
\begin{equation}
U^{(k)} = \begin{bmatrix} V^{(k)} & W^{(k)} \end{bmatrix} \in \mathbb{C}^{m\times m},
\label{eq:unitary_completion}
\end{equation}
where $W^{(k)}\in\mathbb{C}^{m\times(m-n)}$ is chosen such that its columns are orthogonal to those of $V^{(k)}$ and orthonormal themselves, i.e., $(V^{(k)})^\dagger W^{(k)} = 0, (W^{(k)})^\dagger W^{(k)} = I_{m-n}$.
Then $U^{(k)}$ has
% \vskip -9pt
\begin{equation}
\resizebox{\linewidth}{!}{$
\begin{aligned}
&(U^{(k)})^\dagger U^{(k)}
=
\begin{bmatrix}
(V^{(k)})^\dagger \\ (W^{(k)})^\dagger
\end{bmatrix}
\begin{bmatrix}
V^{(k)} & W^{(k)}
\end{bmatrix}
=
\begin{bmatrix}
(V^{(k)})^\dagger V^{(k)} & (V^{(k)})^\dagger W^{(k)}\\
(W^{(k)})^\dagger V^{(k)} & (W^{(k)})^\dagger W^{(k)}
\end{bmatrix}
=
\begin{bmatrix}
I_n & 0\\
0 & I_{m-n}
\end{bmatrix}
= I_m,
\end{aligned}
$}
\end{equation}
and likewise $U^{(k)}(U^{(k)})^\dagger=I_m$.
In practice, $W^{(k)}$ can be obtained by any orthonormal completion procedure (e.g., Gram-Schmidt process \citep{golub2013matrix} used in our experiments). The completion is not unique, and any valid completion suffices for circuit realization.

Finally, the sequential quantum circuit is constructed by applying the local unitaries in order. Each unitary $U^{(k)}$ acts on $(N_b+2)$ qubits, as detailed in \cref{app:unitary_synthesis}, where $N_b=\lceil\log_2 r\rceil$ denotes the number of bond qubits and the two additional qubits correspond to the physical indices $i_k,j_k$; unused bond states are zero-padded when $r$ is not a power of two.
Starting from the initial state $\ket{0}^{\otimes n}$, the full quantum state is prepared by sequentially applying these unitaries, i.e., $\ket{\psi} = U^{(L)}  \cdots U^{(2)} U^{(1)} \ket{0}^{\otimes n}$.

\subsection{Core-wise encoding}
\label{subsubsec:corewise_encoding}

In this strategy, each core tensor is encoded independently into a dedicated subset of qubits, and the resulting registers are combined to form the final quantum circuit, as illustrated in \cref{fig:TNQE}c. Unlike TNQE-full, which aims to prepare a single globally entangled state over all cores, this strategy avoids inter-core entangling operations and thus yields a shallow and modular circuit.

\noindent\textbf{Local encoding of a single core.}
Given a QTT core $\mathcal{A}^{(k)}$,
we flatten it into a vector of length $4r_{k-1}r_k$ and zero-pad it to $\mathbf{v}^{(k)} \in \mathbb{C}^{N_k}$, where $N_k$ is the smallest power of two no smaller than $4r_{k-1}r_k$.
We then normalize it to form
\begin{equation}
\ket{\psi_k}
=
\frac{\mathbf{v}^{(k)}}{\|\mathbf{v}^{(k)}\|_2},
\qquad
\|\mathbf{v}^{(k)}\|_2 > 0.
\label{eq:amp_encode_core}
\end{equation}

Each $\ket{\psi_k}$ is prepared on a dedicated register of  $\log_2(N_k)$ qubits via amplitude encoding \citep{placidi2023mnisq,ran2023tensor}.
Thus, the power-of-two requirement does not impose hard constraints on the ranks used during tensor decomposition.

\noindent\textbf{Quantum circuit construction.}
The full encoded state is obtained by preparing all cores on disjoint qubit registers, $\ket{\Psi} = \bigotimes_{k=1}^{L} \ket{\psi_k}$.
By construction, the circuit introduces no entangling gates across different registers (sub-circuits). As a result, the sub-circuits can be executed independently and in parallel, and the overall circuit depth is bounded by the maximum depth of the individual sub-circuits, rather than scaling with operations on the full $2^{2L}$-dimensional Hilbert space.
We stress that $\ket{\Psi}$ stores the QTT factors rather than the contracted global state $\ket{\psi_X}$; the image is decoded by reading out each register and contracting the bonds classically.

\subsection{Unitary-aware tensor network optimization}
\label{subsec:unitary_aware_qtt}
While TNQE-full is effective, its post hoc unitary synthesis requires an iterative quantum Shannon decomposition (\cref{app:unitary_synthesis}), which can still result in deep circuits.

Here, we establish a distinct paradigm in which TNQE-unitary directly optimizes quantum operations and interprets the resulting unitaries as tensor cores. This design ensures that all intermediate representations correspond to executable quantum operators, eliminates the need for unitary completion after optimization, and enables explicit control over circuit complexity through parameterization.

\begin{wrapfigure}[16]{r}{0.5\textwidth}
\begin{center}
\vskip -13pt
\includegraphics[width=\linewidth]{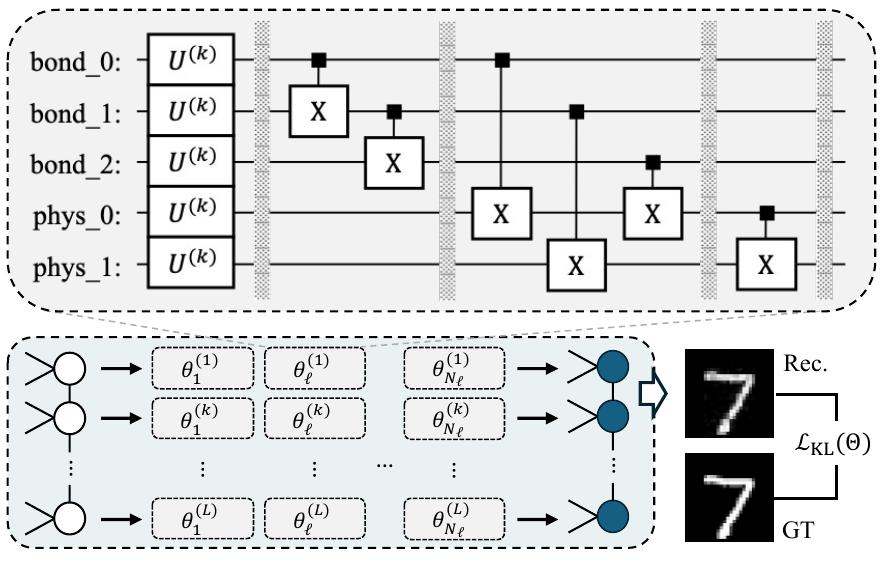}
% \vskip -5pt
\caption{Illustration of TNQE-unitary. Each core is parameterized by a learnable block unitary acting on the bond and physical qubits, with $N_\ell$ layers of single-qubit rotations and fixed entanglers.}
\label{fig:block_unitary}
\end{center}
\end{wrapfigure}

\noindent\textbf{Parameterized block unitaries.}
For each tensor core, we introduce a learnable block unitary
$
U^{(k)}(\theta^{(k)}),
\theta^{(k)} \in \mathbb{R}^{(2 \times (N_b+2) \times N_\ell)},
$
where $\theta$ denotes all optimized parameters, including two rotation gates on each of the $N_b+2$ qubits across $N_\ell$ circuit layers.
As detailed in \cref{fig:block_unitary}, we implement $U^{(k)}(\theta^{(k)})$ using (a) single-qubit rotations and (b) fixed entangling gates in each of the $N_\ell$ layers:

a)
For layer $\ell$ and qubit $q$, we have
\begin{equation}
U_{1q}(\alpha_{\ell,q}^{(k)},\beta_{\ell,q}^{(k)})
=
R_Z(\beta_{\ell,q}^{(k)})\,R_Y(\alpha_{\ell,q}^{(k)}),
\end{equation}
with \citep{nielsen2010quantum}
\begin{equation}
  \begin{aligned}
R_Y(\alpha) = e^{-i\alpha Y/2}
=
\begin{pmatrix}
\cos(\alpha/2) & -\sin(\alpha/2)\\
\sin(\alpha/2) & \cos(\alpha/2)
\end{pmatrix},
R_Z(\beta) = e^{-i\beta Z/2}
=
\begin{pmatrix}
e^{-i\beta/2} & 0\\
0 & e^{i\beta/2}
\end{pmatrix}.
\label{eq:ry_rz}
  \end{aligned}
\end{equation}

b) Following hardware-efficient designs \citep{kandala2017hardware,huggins2019towards}, the entangler $U_e$ is composed of CNOT gates with a fixed pattern including bond-to-bond $\mathrm{CNOT}_{q \rightarrow q+1}$, bond-to-physical $\mathrm{CNOT}_{q \rightarrow \mathrm{p}(q)}$, and physical-to-physical $\mathrm{CNOT}_{\mathrm{p_0}\rightarrow \mathrm{p_1}}$.

Consequently, the full block unitary is constructed by
\begin{equation}
\resizebox{\linewidth}{!}{$
\begin{aligned}
U^{(k)}(\theta^{(k)})
=
\prod_{\ell=1}^{N_\ell}
(U_e \, U_{\ell}^{(k)}),
U_{\ell}^{(k)}
=
\bigotimes_{q=0}^{n-1}
U_{1q}(\alpha_{\ell,q}^{(k)},\beta_{\ell,q}^{(k)}),
\label{eq:block_unitary}
\end{aligned}
$}
\end{equation}
as illustrated in \cref{fig:block_unitary}.

\noindent\textbf{From unitaries to isometric QTT cores.}
To recover an isometry compatible with QTT cores, we extract the first $n$ columns of $U^{(k)}$, i.e.,
$V^{(k)} = U^{(k)}(:,0{:}n) \in \mathbb{C}^{m \times n}$.
We then reshape $V^{(k)}$ into the QTT cores $\mathcal{A}^{(k)}$ by interpreting its row index as a composite of physical outputs $i_k,j_k$ and bond output $r_k$:
\begin{equation}
\mathcal{A}^{(k)}_{r_{k-1},i_k,j_k,r_k} \equiv \bra{r_k,i_k,j_k} V^{(k)} \ket{r_{k-1}},
\end{equation}
with the boundary conditions $r_0=r_L=1$ imposed by fixing the bond input of the first block and the bond output of the last block to the $\ket{0}$ index. In this way, the tensor network is defined by a set of quantum operations, and the resulting cores are directly interpretable as executable quantum circuits.

\noindent\textbf{Optimization objective.}
Let $X(\Theta)$ denote the reconstructed image obtained by contracting the resulting QTT cores, where $\Theta=\{\theta^{(k)}\}_{k=1}^L$.
We optimize $\Theta$ by minimizing the Kullback–Leibler divergence \citep{kullback1951information} between the normalized target image $Y$ and the Born probability distribution induced by $X(\Theta)$:
\begin{equation}
\mathcal{L}_{\mathrm{KL}}(\Theta)
=
\sum_{i,j}
p_Y(i,j)\,
\Bigl[
\log p_Y(i,j)
-
\log p_X(i,j;\Theta)
\Bigr],
\label{eq:kl_loss}
\end{equation}
where
$p_X(i,j;\Theta)=|X_{ij}(\Theta)|^2/\sum_{i,j}|X_{ij}(\Theta)|^2$ and
$p_Y(i,j)=Y_{ij}/\sum_{i,j}Y_{ij}$.
After optimization, the learned parameters $\Theta$ immediately specify the corresponding quantum circuit blocks $U^{(k)}(\theta^{(k)})$ without additional unitary completion or gate synthesis. Since \cref{eq:kl_loss} matches Born probabilities, TNQE-unitary follows the probability convention of \cref{sec:method} ($|X_{ij}|\propto\sqrt{Y_{ij}}$); the decoding rule and the relation to the amplitude convention are detailed in \cref{app:semantics}.

\textbf{Remark.}
Amplitude encoding prepares a global state over the entire pixel space with operations acting across many qubits (\cref{fig:circuits}a), which leads to deep circuits in practice. TNQE instead operates on local tensor cores: TNQE-full and TNQE-unitary map each core to a unitary on $N_b+2$ qubits (\cref{fig:circuits}b), and TNQE-core encodes cores on disjoint sub-circuits without long-range entangling operations (\cref{fig:circuits}c). Overall, by operating on tensor cores rather than directly on the global pixel space, TNQE provides a structured and hardware-compatible strategy to quantum data encoding.

\section{Experiments}
\label{sec:experiments}
\subsection{Experimental setup}
\noindent\textbf{Dataset:} We evaluate the proposed TNQE methods on the widely used MNIST dataset \citep{lecun1998mnist}, which represents a challenging benchmark for quantum data encoding. MNIST consists of handwritten digits with a resolution of $28 \times 28$ pixels. To accommodate the power-of-two structure required by our tensor network representation, all images are zero-padded with a padding width of 2, resulting in images of size $32 \times 32$. For TNQE-unitary, we experiment on a subset of 1000 MNIST images to explore the impact of different hyperparameters on performance. For the $256 \times 256$ experiments, we use two well-known images: the grayscale image Cameraman and the color image Girl with a Pearl Earring.

\noindent\textbf{Evaluation metrics:}
We evaluate the proposed TNQE methods using multiple criteria, including image approximation quality and circuit structure. Image approximation quality is measured using mean squared error (MSE), binary cross entropy (BCE), peak signal-to-noise ratio (PSNR), and structural similarity index measure (SSIM). We further evaluate the semantic information preserved in these images using two classical classifiers.

\noindent\textbf{Implementation details:} We use Qiskit \citep{javadi2024quantum} to simulate the quantum circuits generated by the proposed TNQEs. The simulations are performed on a classical computer with sufficient computational resources to handle the quantum circuit simulations.

\subsection{Quantum circuit comparison}

\begin{figure}[htbp]
\centering
\begin{minipage}[t]{0.65\textwidth}
\vspace{0pt}
\centering
\captionof{table}{Comparison of quantum circuits on $32\times32$ images.}
\label{tab:circuit_comparison}
\resizebox{\linewidth}{!}{
\begin{tabular}{lccccc}
\toprule
Methods & Qubits & Depth & Unitary & CNOT & Operators \\
\midrule
Baseline (Amplitude) & \textbf{10}    & 2028     & 1023 & 1013 & 2036 \\
Automatic encoding & \textbf{10}    & 208   & 371 & 129 & 500 \\
TNQE-full & 12    & 945   & 900 & 475 & 1375 \\
TNQE-core & 26    & 116   & 219 & 193 & 412 \\
TNQE-unitary & 13 & \textbf{81} & \textbf{100} & \textbf{120} & \textbf{220} \\
\bottomrule
\label{tab:mnist_com}
\end{tabular}}
\end{minipage}
\hfill
\begin{minipage}[t]{0.31\linewidth}
\vspace{0pt}
\centering
\includegraphics[width=\linewidth]{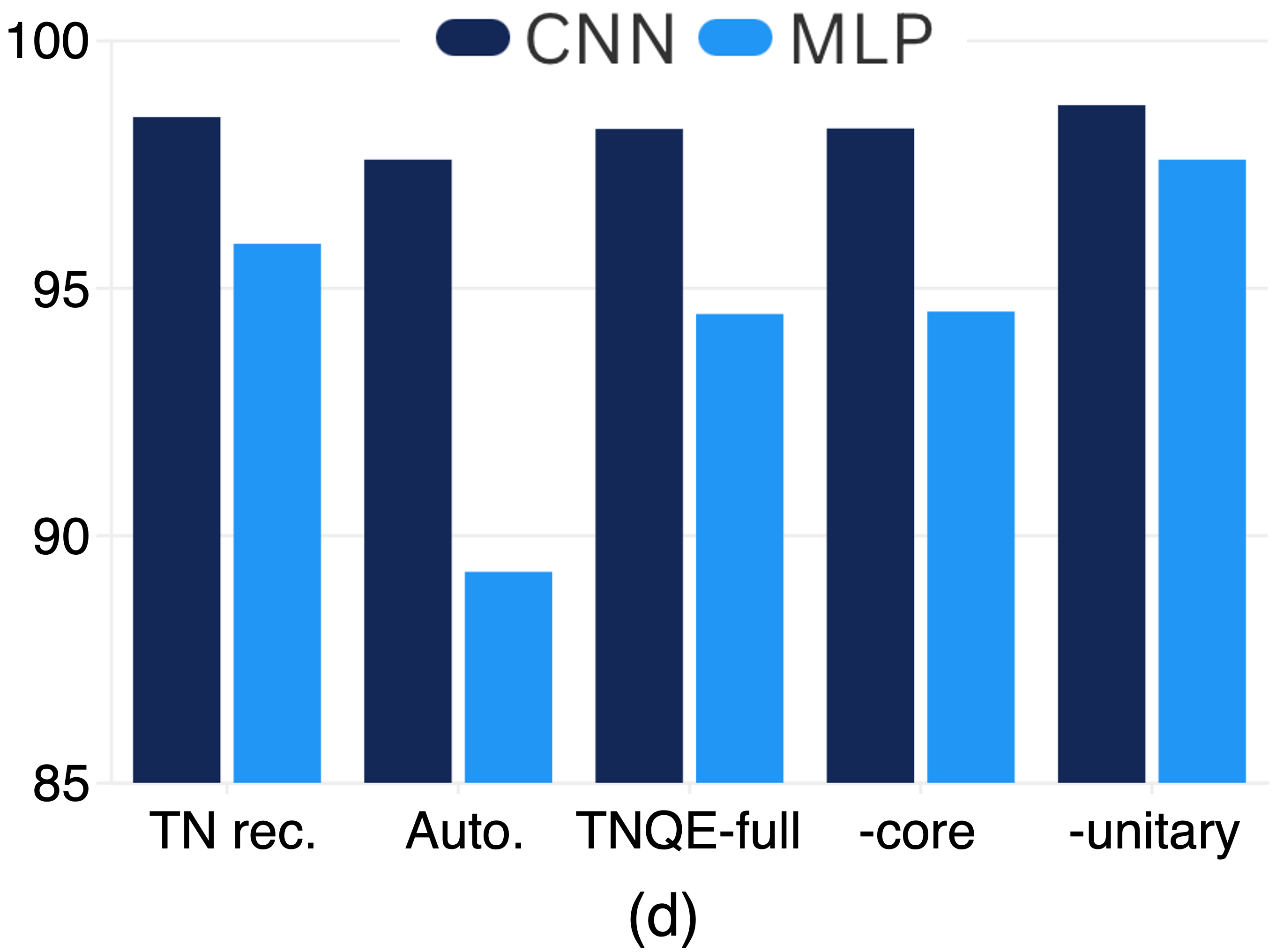}
\end{minipage}
\includegraphics[width=\textwidth]{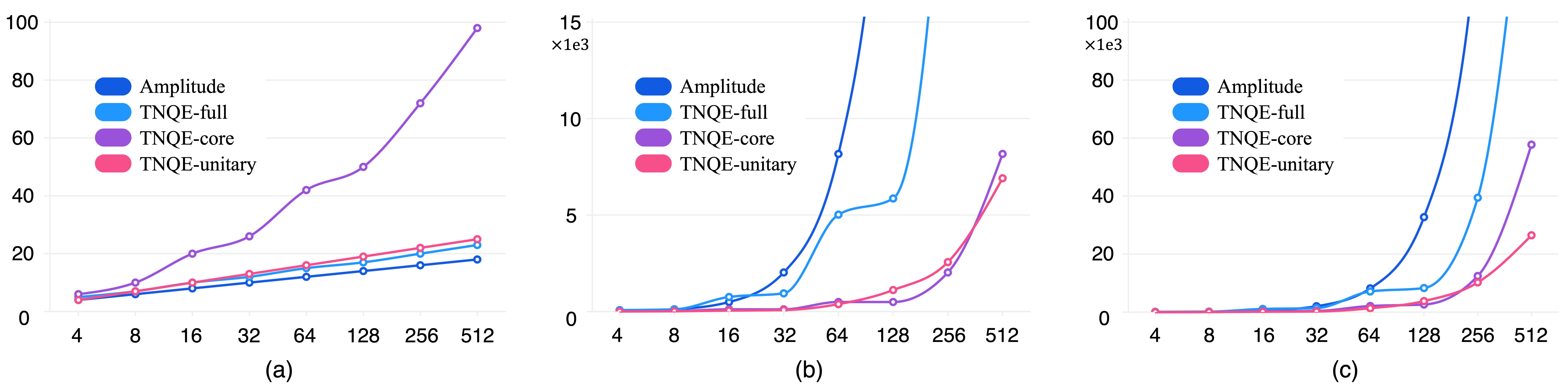}
\caption{Scalability of different quantum data encoding methods. (a) The number of qubits, (b) circuit depth, and (c) the number of quantum operators as image size increases from $4 \times 4$ to $512 \times 512$. (d) Classification accuracy (\%) on approximated MNIST images using CNN and MLP classifiers.}
\label{fig:circuit_com}
\end{figure}

In this section, we construct quantum circuits for MNIST images and compare the circuit structure of multiple methods, as summarized in \cref{tab:mnist_com}.
We set $r = 4$ for TNQE-full and TNQE-core, and $r = 8$ with $N_\ell = 4$ for TNQE-unitary.

\renewcommand{\arraystretch}{0.8}

As discussed earlier, basis encoding requires 1024 qubits, making it impractical for current quantum hardware. Amplitude encoding \citep{schuld2019quantum} reduces the qubit requirement to 10 by encoding the entire pixel space into a quantum circuit, but at the cost of a deep circuit with a large number of operators. We refer to this method as a key baseline for comparison.
\citet{placidi2023mnisq} adopt automatic encoding \citep{shirakawa2024automatic} to construct shallower and more efficient circuits for MNIST images; however, this also introduces noticeable approximation errors, as shown in \cref{fig:mnist_visual}.
In contrast, TNQEs leverage tensor network representations rather than entire pixels to achieve a more favorable balance between qubit count and circuit depth.
The TNQE-full strategy introduces two additional bond qubits, yet results in a $0.47\times$ circuit depth compared to the baseline.
The TNQE-core strategy further achieves a reduction to only 116 circuit depth ($0.06\times$) by increasing the qubits to 26.
Regarding TNQE-unitary strategy, the circuit depth is controllable and directly determined by setted $N_\ell$. We find that a circuit with only 81 circuit depth ($0.04\times$) is already sufficient to encode MNIST images, which is substantially shallower than that required by all other methods.

In addition, we evaluate the scalability of different quantum data encoding methods as the image size increases from $4 \times 4$ to $512 \times 512$, as shown in \cref{fig:circuit_com}a-c. Amplitude encoding exhibits rapid growth in depth and operators, rendering it impractical for larger images. Compared with this baseline, the advantages of our TNQE methods are substantial. Specifically, TNQE-full effectively reduces circuit complexity while maintaining a small number of qubits. In contrast, TNQE-core leverages a larger number of qubits to achieve a more pronounced reduction in circuit complexity. These two strategies exhibit complementary strengths and are suitable for different application scenarios.
Finally, TNQE-unitary demonstrates the best scalability, with both qubit count and circuit complexity increasing much more slowly with image size than others. This makes TNQE-unitary particularly promising for encoding high-resolution images; note that these curves are at fixed $r$ and $N_\ell$, and the error attained at a fixed budget depends on the compressibility of the input (\cref{app:discussion}).

\subsection{Image approximation quality}

After discussing the circuit complexity, we further evaluate their performance in terms of image approximation quality, including visual and quantitative comparisons.

\begin{figure}[htbp]
\centering
\begin{minipage}[t]{0.48\linewidth}
\vspace{2pt}
\centering
\includegraphics[width=\linewidth]{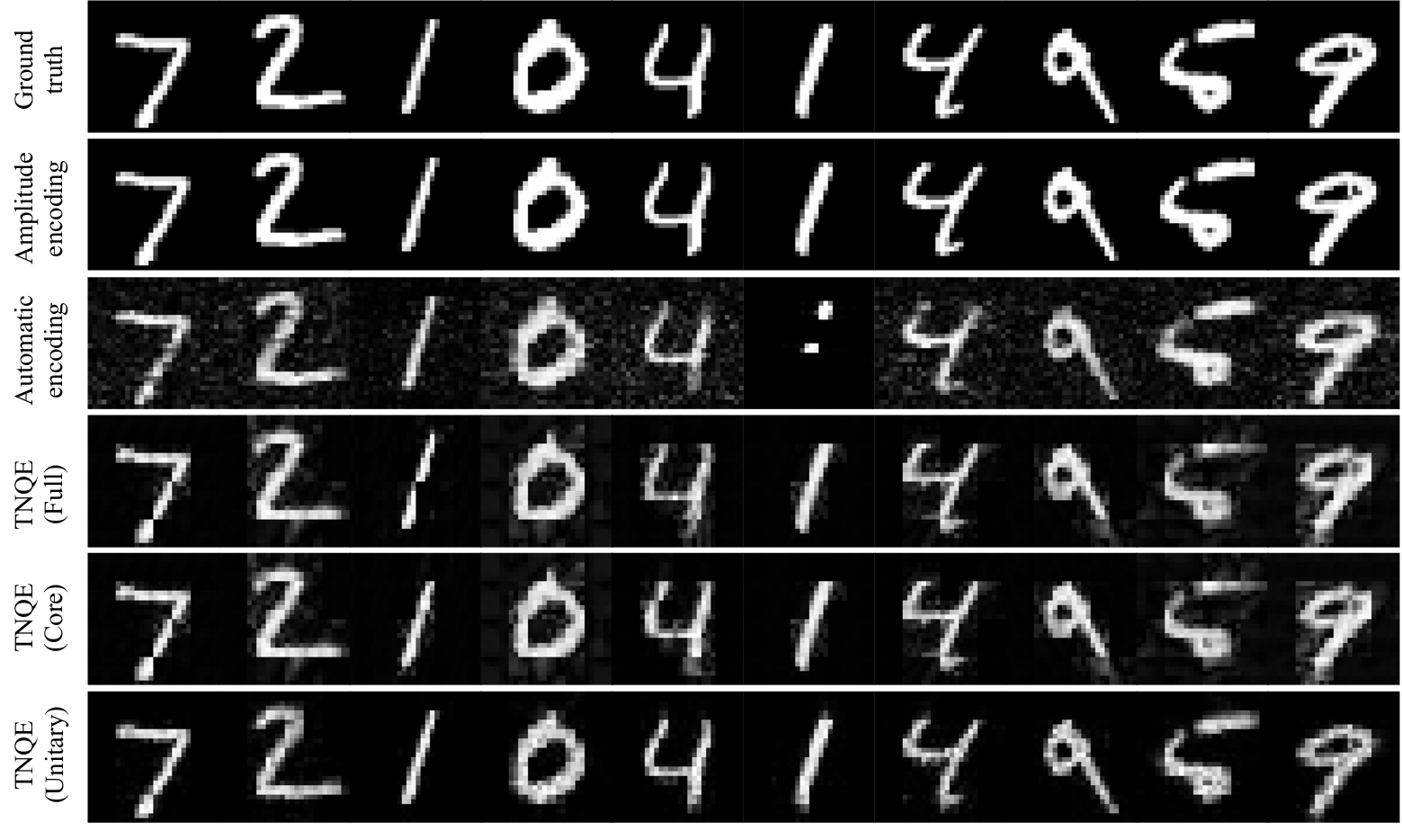}
\caption{Comparison of MNIST images from different encoding methods in simulation.}
\label{fig:mnist_visual}
\end{minipage}\hfill
\begin{minipage}[t]{0.49\textwidth}
\vspace{0pt}
\centering
\captionof{table}{Quantitative comparison of images approximated by quantum data encoding methods in simulation on MNIST.}
\resizebox{\linewidth}{!}{
\begin{tabular}{lcccc}
    \toprule
    Methods & MSE   & BCE   & PSNR & SSIM \\
    \midrule
    \rowcolor{gray!15}
    \multicolumn{5}{l}{\textbf{Results with original image}} \\
    \midrule
    TN reconstruction & 0.013135 & 0.113215 & 19.3358 & 0.924981 \\
    Automatic encoding & 0.015656 & 0.147523 & 18.3669 &  0.882500 \\
    TNQE-full & 0.015700 & 0.129583 & 18.6845 &   0.906974 \\
    TNQE-core & 0.015639 & 0.129311 & 18.6903 & 0.907203 \\
    TNQE-unitary & \textbf{0.011249} & \textbf{0.092333} & \textbf{20.5837} &   \textbf{0.915903} \\
    \midrule
    \rowcolor{gray!15}
    \multicolumn{5}{l}{\textbf{Results with referenced image}} \\
    % \toprule
    % Methods & MSE   & BCE   & PSNR & SSIM \\
    \midrule
    TN reconstruction & 0.013135 & 0.113215 & 19.3358 & 0.924981 \\
    Automatic encoding  & 0.015656 & 0.147523 & 18.3669 & 0.882500 \\
    TNQE-full & 0.002234 & 0.165439 & 28.4850 &   0.983236 \\
    TNQE-core & \textbf{0.002182} & 0.165311 & \textbf{28.4913} & \textbf{0.983443} \\
    TNQE-unitary & 0.011249 & \textbf{0.092333} & 20.5837 &   0.915903 \\
    \bottomrule
    \end{tabular}}
  \label{tab:quantitative_com}
\end{minipage}
\end{figure}

\textbf{Visual comparisons:}
In \cref{fig:mnist_visual}, we present a visual comparison of MNIST images approximated using different quantum data encoding methods in simulation.
The first row displays the original images as ground truth and the second row shows the images from amplitude encoding, which serves as a key baseline for comparison. Automatic encoding \citep{placidi2023mnisq} also encodes the entire pixel space into the amplitudes, which leads to substantial error accumulation and visible artifacts, as shown in the third row. There are pronounced background grain and information loss, which substantially degrade approximation quality.
In contrast, our TNQE strategies encode the tensor network representation rather than operating directly on the entire pixel space. By first factorizing each image into a sequence of tensor cores and then encoding these cores into a quantum circuit, the encoding procedure becomes more local and accurate. Notably, each step acts on a single tensor core and applies only local operations on a small subset of qubits, instead of performing complex global operations across all qubits, keeping the circuits shallow while preserving the digit structure.
% This pipeline significantly reduces circuit complexity and error accumulation, demonstrating that TNQE provides a promising solution for quantum data encoding.

\textbf{Quantitative comparisons:}
We compare different quantum data encoding methods on MNIST using multiple evaluation metrics, including MSE, BCE, PSNR, and SSIM. In addition, we evaluate classification accuracy on approximated images as an indicator of semantic information preservation.

As shown in the top part of \cref{tab:quantitative_com}, TNQE-unitary consistently outperforms other circuit encodings across all metrics, demonstrating its ability to preserve information with fewer quantum operators. When evaluated with their respective reference images, the other two TNQE methods also show clear advantages. Specifically, original images serve as references for TN, automatic, and TNQE-unitary encodings, while TN-reconstructed images are used for the remaining two TNQE methods. Under this comparison, TNQE-core achieves an MSE of 0.002182, a PSNR of 28.4913, and an SSIM of 0.983443.
These results suggest that advances in TNs can improve TNQE, pointing to a promising direction in the future. We further provide a more detailed analysis of the TNQE pipeline in the appendix, covering the decomposition and quantum data encoding processes (see \cref{tab_app:org_com,tab_app:stage_com}).

For a more comprehensive evaluation, we conduct classification experiments using CNN and MLP on MNIST, as shown in \cref{fig:circuit_com}d and \cref{fig:confusion_matrix}. The TNQEs maintain high accuracy, and similarly to the above observations, TNQE-unitary achieves the best accuracy while using fewer quantum operators. This demonstrates that the TNQE methods effectively preserve the semantic information of images, and that directly optimizing the quantum operators is an efficient strategy.

\subsection{Ablation and real quantum experiments}
\label{subsec:ablation_hw}
\begin{figure}[htbp]
\centering
\begin{minipage}[t]{0.49\linewidth}
\vspace{7pt}
\begin{center}
\includegraphics[width=\linewidth]{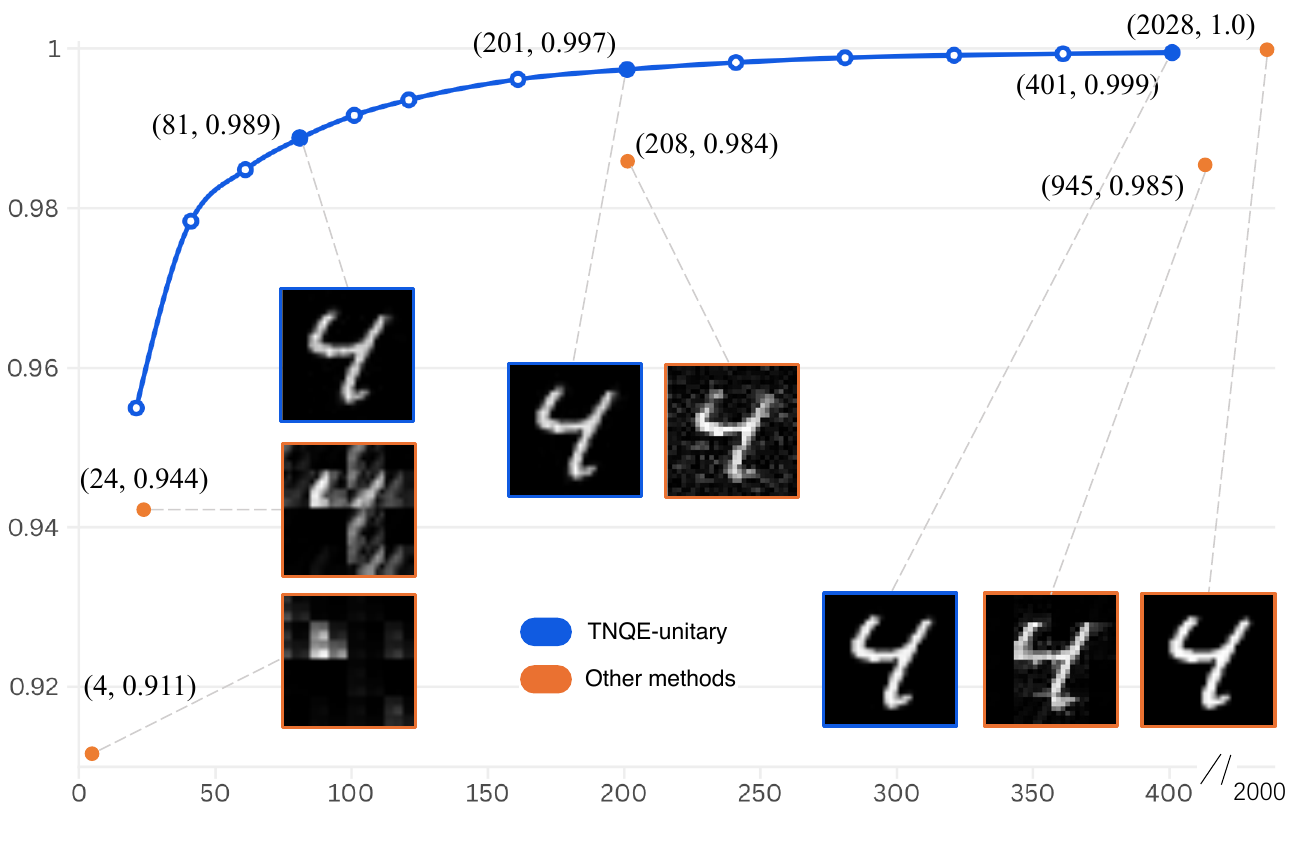}
\caption{Impact of $N_\ell$ on TNQE-unitary. For comparison with other methods, the x-axis reports the corresponding circuit depth and the y-axis reports $1-\text{MSE}$. Detailed numerical results are provided in \cref{tab:per_depth}.}
\label{fig:per_depth}
\end{center}
\end{minipage}\hfill
\begin{minipage}[t]{0.48\textwidth}
\vspace{-4pt}
  \centering
  \captionof{table}{Comparisons on real quantum hardware.}
  \resizebox{\linewidth}{!}{
    \begin{tabular}{cccccc}
    \toprule
    QPUs  & Methods & MSE   & BCE   & PSNR  & SSIM \\
    \midrule
    \multirow{3}{*}{\begin{tabular}[c]{@{}c@{}}ibm\_torino\\(Heron r1)\end{tabular}} & Amp.  & 0.1122 & 0.4880 & 9.6688 & 0.0084 \\
          & Auto. & 0.0508 & 0.2682 & 13.2173 & 0.5545 \\
          & Ours  & \textbf{0.0342} & \textbf{0.1930} & \textbf{15.0322} & \textbf{0.7357} \\
    \midrule
    \multirow{3}{*}{\begin{tabular}[c]{@{}c@{}}ibm\_fez\\(Heron r2)\end{tabular}} & Amp.  & 0.1077 & 0.6667 & 9.9193 & 0.0003 \\
          & Auto. & 0.0531 & 0.2708 & 13.2744 & 0.5674 \\
          & Ours  & \textbf{0.0299} & \textbf{0.1833} & \textbf{15.6425} & \textbf{0.7737} \\
    \midrule
    \multirow{3}{*}{\begin{tabular}[c]{@{}c@{}}ibm\_boston\\(Heron r3)\end{tabular}} & Amp.  & 0.1075 & 0.5686 & 9.9130 & 0.0225 \\
          & Auto. & 0.0418 & 0.2210 & 14.2244 & 0.6278 \\
          & Ours & \textbf{0.0283} & \textbf{0.1748} & \textbf{15.8837} & \textbf{0.7894} \\
    \bottomrule
    \end{tabular}}
  \label{tab:real_quantum}
\begin{center}
\vskip 2pt
\includegraphics[width=\linewidth]{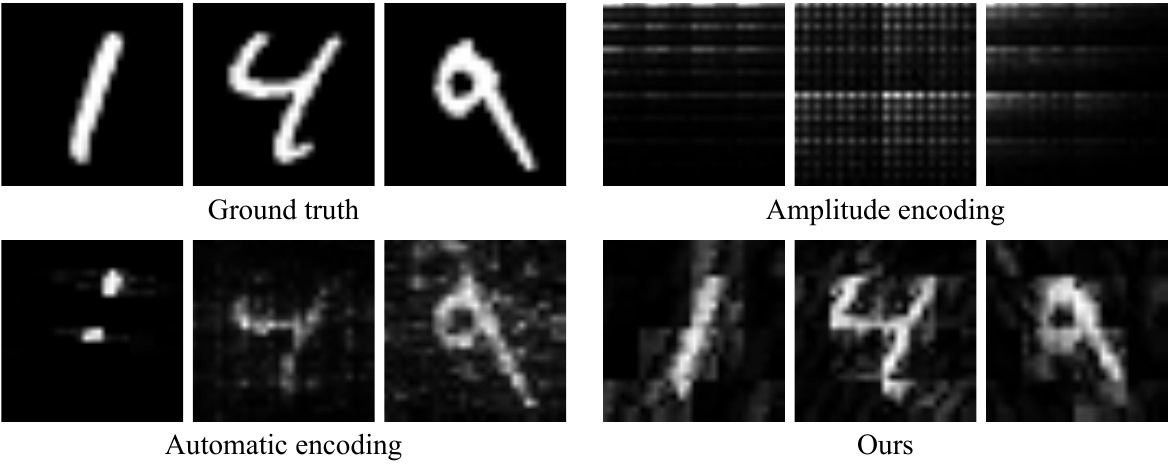}
\captionof{figure}{Visual comparisons on ibm\_boston.}
\label{fig:real_quantum}
\end{center}
\end{minipage}
% \vskip -9pt
\end{figure}

In this section, we conduct ablation studies to analyze the impact of key hyperparameters with 1K MNIST images.
As shown in \cref{tab:ablation_mse} and \cref{tab:ablation_others}, we vary the tensor network rank and evaluate image approximation quality. Since the number of bond qubits depends on $\lceil\log_2(r)\rceil$, we conduct experiments over ranks corresponding to 0 to 4 bond qubits. For the first two TNQE methods, increasing the rank generally improves performance because larger ranks yield more expressive tensor network representations; however, this improvement comes at the cost of a substantial increase in circuit depth. For TNQE-unitary, depth is primarily governed by $N_\ell$ and changes only when the number of bond qubits increases, as shown in \cref{fig:per_depth}.

{\renewcommand{\arraystretch}{1.}
\begin{table}[t]
  \centering
  \caption{Ablation study of different ranks using TNQE on MNIST. The circuit structure is reported as (qubits, depth).}
  \vskip 9pt
  \resizebox{\linewidth}{!}{
    \begin{tabular}{cccccccccc}
    \toprule
    TNQEs & Rank & 1     & 2     & 3     & 4     & 5     & 8     & 9     & 16 \\
    \midrule
    \multirow{2}{*}{Full} & MSE   & 0.0889 & 0.0562 & 0.0321 & 0.0147 & 0.0089 & 0.0018 & 0.0012 & 0.0030 \\
          & Circuit & (10, 7) & (11, 195) & (12, 945) & (12, 945) & (13, 4185) & (13, 4185) & (14, 17625) & (14, 17625) \\
    \midrule
    \multirow{2}{*}{Core} & MSE   & 0.0891 & 0.0562 & 0.0322 & 0.0147 & 0.0089 & 0.0019 & 0.0012 & 0.0030 \\
          & Circuit & (10, 4) & (18, 24) & (26, 116) & (26, 116) & (31, 242) & (34, 496) & (39, 1006) & (42, 2028) \\
    \midrule
    \multirow{2}{*}{Unitary} & MSE   & 0.1626 & 0.0450 & 0.0184 & 0.0185 & 0.011287 & 0.0112 & 0.0084 & 0.0083 \\
          & Circuit & (10, 8) & (11, 56) & (12, 61) & (12, 61) & (13, 81) & (13, 81) & (14, 82) & (14, 82) \\
    \bottomrule
    \end{tabular}}
  \label{tab:ablation_mse}
  % \vskip -5pt
\end{table}}

Similarly, increasing $N_\ell$ enhances the expressiveness of the block unitary, leading to increased circuit depth and improved image approximation quality. We also present the performance of other methods at different depths. TNQE-unitary achieves strong performance at relatively small depths, highlighting its high encoding efficiency.

\textbf{Experiment on real quantum hardware:}
We further validate TNQE-core on three quantum processing units (QPUs) using 100 MNIST images, as shown in \cref{tab:real_quantum}. The depths reported in this paper are logical depths in a gate set of single-qubit rotations and CNOTs before device-specific routing, and no error mitigation or hardware-specific optimization is applied.
While deeper circuits provide stronger representational capacity, as exemplified by amplitude encoding in \cref{fig:mnist_visual}, hardware noise causes such circuits to suffer from catastrophic error accumulation, as shown in \cref{fig:real_quantum}. In contrast, our method preserves semantic information under hardware noise. We regard this as initial evidence of hardware feasibility rather than a demonstration of hardware-level advantage; additional discussion and a high-resolution stress test are provided in \cref{app:discussion}.

\section{Conclusion}
In this work, we propose TNQE, a TN-based framework for circuit-efficient quantum data encoding. By leveraging the intrinsic connections between TNs and quantum systems, TNQE establishes a practical pipeline for transforming classical data into executable quantum circuits through structured conversion.
Extensive experiments demonstrate that TNQE enables shallow and resource-efficient quantum circuits while preserving informative structure. Notably, TNQE attains lower error than approximate baselines at circuit depths as low as $0.04\times$ that of exact amplitude encoding, naturally scales to high-resolution data, and exhibits encouraging practicality on real quantum hardware
Together, these results highlight TNQE as a promising framework for quantum data encoding and motivate several directions for further extending TNQE, as discussed in \cref{app:discussion}.

\section*{Acknowledgments}
Guang Lin was supported by the RIKEN Special Postdoctoral Researchers Program.
This work was partially supported by the RIKEN TRIP initiative (RIKEN Quantum), UTokyo Quantum Initiative, JSPS KAKENHI Grant Number JP23K28109, and JSPS Bilateral Program Number JPJSBP120257420.

\bibliography{neurips_2026}
\bibliographystyle{neurips_2026}

%%%%%%%%%%%%%%%%%%%%%%%%%%%%%%%%%%%%%%%%%%%%%%%%%%%%%%%%%%%%

% \newpage
\appendix
\newpage
\section{Right-canonicalization by SVD}
\label{app:right_canonicalization}

In this section, we provide a detailed explanation of the right-canonicalization procedure used in \cref{subsubsec:full_tn_encoding}, in particular clarifying the meaning of absorbing the non-isometric factor into the neighboring tensor to the left.

Consider a three-way MPS tensor at core $k$,
\begin{equation}
A^{(k)} \in \mathbb{C}^{r_{k-1} \times d \times r_k},
\end{equation}
where $r_{k-1}$ and $r_k$ denote the left and right bond dimensions (ranks), respectively, and $d$ is the physical dimension (with $d=4$ in our construction).
To perform right-canonicalization, we reshape $A^{(k)}$ into a matrix by grouping the physical index and the right bond index,
\begin{equation}
\mathbf{A}^{(k)} \in \mathbb{C}^{r_{k-1} \times (d r_k)},
\qquad
\mathbf{A}^{(k)}_{\alpha_{k-1},(p_k,\alpha_k)} = A^{(k)}_{\alpha_{k-1},p_k,\alpha_k}.
\label{eq:appendix_reshape}
\end{equation}

Then,
we compute the singular value decomposition (SVD)
\begin{equation}
\mathbf{A}^{(k)} = W^{(k)} \Sigma^{(k)} \bigl(V^{(k)}\bigr)^\dagger,
\label{eq:appendix_svd}
\end{equation}
where $W^{(k)} \in \mathbb{C}^{r_{k-1} \times r_{k-1}}$ and
$V^{(k)} \in \mathbb{C}^{(d r_k) \times r_{k-1}}$ have orthonormal columns, and
$\Sigma^{(k)}$ is a non-negative diagonal matrix containing all singular values; no truncation is performed during canonicalization.
Since the columns of $V^{(k)}$ are orthonormal, the matrix
$\bigl(V^{(k)}\bigr)^\dagger \in \mathbb{C}^{r_{k-1} \times (d r_k)}$
satisfies the row-isometric condition
\begin{equation}
\bigl(V^{(k)}\bigr)^\dagger \bigl((V^{(k)})^\dagger\bigr)^\dagger
=
\bigl(V^{(k)}\bigr)^\dagger V^{(k)}
=
I_{r_{k-1}}.
\end{equation}
Therefore, $\bigl(V^{(k)}\bigr)^\dagger$ can be directly reshaped into a right-canonical MPS tensor at core $k$.
The remaining factor $W^{(k)} \Sigma^{(k)}$ is generally not isometric and hence cannot be retained at core $k$.
Instead, this factor is multiplied into the neighboring matrix at core $k-1$ by contracting it with the corresponding bond index:
\begin{equation}
\mathbf{A}^{(k-1)} \;\leftarrow\; \mathbf{A}^{(k-1)} \cdot \bigl(W^{(k)} \Sigma^{(k)}\bigr).
\label{eq:appendix_absorb}
\end{equation}
This operation preserves the value of the overall tensor network, since it merely redistributes factors within a local product of matrices, while ensuring that the updated matrix at core $k$ remains right-canonical.
Before the multiplication, $\mathbf{A}^{(k-1)}$ is reshaped into a matrix of shape $(r_{k-2}d)\times r_{k-1}$, such that the factor $W^{(k)}\Sigma^{(k)} \in \mathbb{C}^{r_{k-1}\times r_{k-1}}$ can be absorbed through standard matrix multiplication. Equivalently, this operation can be interpreted as a tensor contraction along the shared bond index $r_{k-1}$.
By iterating the above steps from $k=L$ down to $k=2$, all matrices except the first are transformed into right-canonical form without changing the represented tensor.
As a consequence, the norm of the global MPS depends solely on the first matrix $\mathbf{A}^{(1)}$, allowing the global normalization $\|\psi\|=1$ to be enforced by rescaling $\mathbf{A}^{(1)}$ alone.

\section{Synthesis of unitaries into elementary quantum gates}
\label{app:unitary_synthesis}
After obtaining the local unitaries, we adopt Quantum Shannon Decomposition \citep[QSD,][]{shende2005synthesis} based on the Block ZXZ-Decomposition \citep{krol2024beyond} to decompose each unitary into elementary quantum gates, such as single-qubit rotations and CNOT gates. Specifically, given an $n$-qubit unitary $U \in \mathbb{C}^{2^n \times 2^n}$, QSD expresses $U$ recursively as a product of $(n-1)$-qubit unitaries and single-qubit rotations. By partitioning $U$ with respect to a selected control qubit, $U$ can be written in block form as
\begin{equation}
U =
\begin{pmatrix}
U_{00} & U_{01} \\
U_{10} & U_{11}
\end{pmatrix},
\qquad
U_{ij} \in \mathbb{C}^{2^{n-1} \times 2^{n-1}}.
\end{equation}

In the iterative QSD procedure, the same factorization is applied successively to the resulting $(n-1)$-qubit unitaries until reaching base cases of one- or two-qubit unitaries, which can be synthesized directly. This synthesis process is implemented in the UnitaryGate module of Qiskit \citep{javadi2024quantum}, which we utilize to convert the local unitaries into quantum circuits.

\section{Target-state semantics of TNQE-unitary}
\label{app:semantics}

Since \cref{eq:kl_loss} matches Born probabilities to $p_Y$, the optimized state has $|X_{ij}|\propto\sqrt{Y_{ij}}$: TNQE-unitary follows the probability convention, and its decoded image is $\hat{Y}_{ij}=p_X(i,j;\Theta)\sum_{i,j}Y_{ij}$, whereas amplitude encoding, automatic encoding, TNQE-full, and TNQE-core follow the amplitude convention $X_{ij}\propto Y_{ij}$. The two conventions therefore prepare different states for the same image (amplitudes $\propto\sqrt{Y}$ versus $\propto Y$), and all comparisons in the main text are made in pixel space between decoded images and the original.

The probability convention is a design choice rather than a restriction: replacing $p_Y$ by $Y_{ij}^2/\sum_{i,j}Y_{ij}^2$ in \cref{eq:kl_loss} targets the amplitude-convention state instead. We adopt it because the image is then decodable from measurement counts alone, without sign or phase information. Finally, unlike variational state preparation, which fits a circuit to a target state through simulated circuit evaluations, the objective of TNQE-unitary is a classical QTT contraction with no simulated circuit execution or shot-based estimation, and the circuit layout ($L$ blocks on $N_b+2$ qubits with $N_\ell$ layers each) is fixed before training.

\section{Introducing rotations for circuit parameterization}

In TNQE-unitary, the goal of the parameterized quantum circuit is to realize a family of unitaries that can be directly mapped onto quantum circuits. We therefore adopt a gate-based parameterization strategy, choosing $R_Y(\alpha)$ and $R_Z(\alpha)$ gates and together with entangling gates.

For a single qubit, any unitary operator $U \in \mathbb{C}^{2 \times 2}$ can be expressed using three continuous rotation parameters,  given by the Euler-angle decomposition
\begin{equation}
U(\gamma, \alpha,\beta)
=
R_X(\gamma) R_Y(\alpha) R_Z(\beta),
\label{eq:euler_xyz}
\end{equation}
where the rotation operators are defined as
\begin{equation}
R_\mu(\theta) = e^{-i\theta \mu/2}
\qquad \mu \in \{X,Y,Z\}.
\end{equation}

Although \Cref{eq:euler_xyz} explicitly involves rotations around all three Pauli axes, the $R_X$ rotation does not add expressiveness. In particular, an $X$-axis rotation can be exactly represented using only $Y$- and $Z$-axis rotations via
\begin{equation}
R_X(\theta)
=
R_Z(-\frac{\pi}{2})
R_Y(\theta)
R_Z(\frac{\pi}{2}).
\end{equation}

This shows that $R_Y$ and $R_Z$ are sufficient to parameterize arbitrary single-qubit unitaries in TNQE-unitary. Including an explicit $R_X$ gate would therefore be redundant. In this work, we restrict the single-qubit gate set to $R_Y$ and $R_Z$.

\begin{figure}[htbp]
\begin{center}
\includegraphics[width=\linewidth]{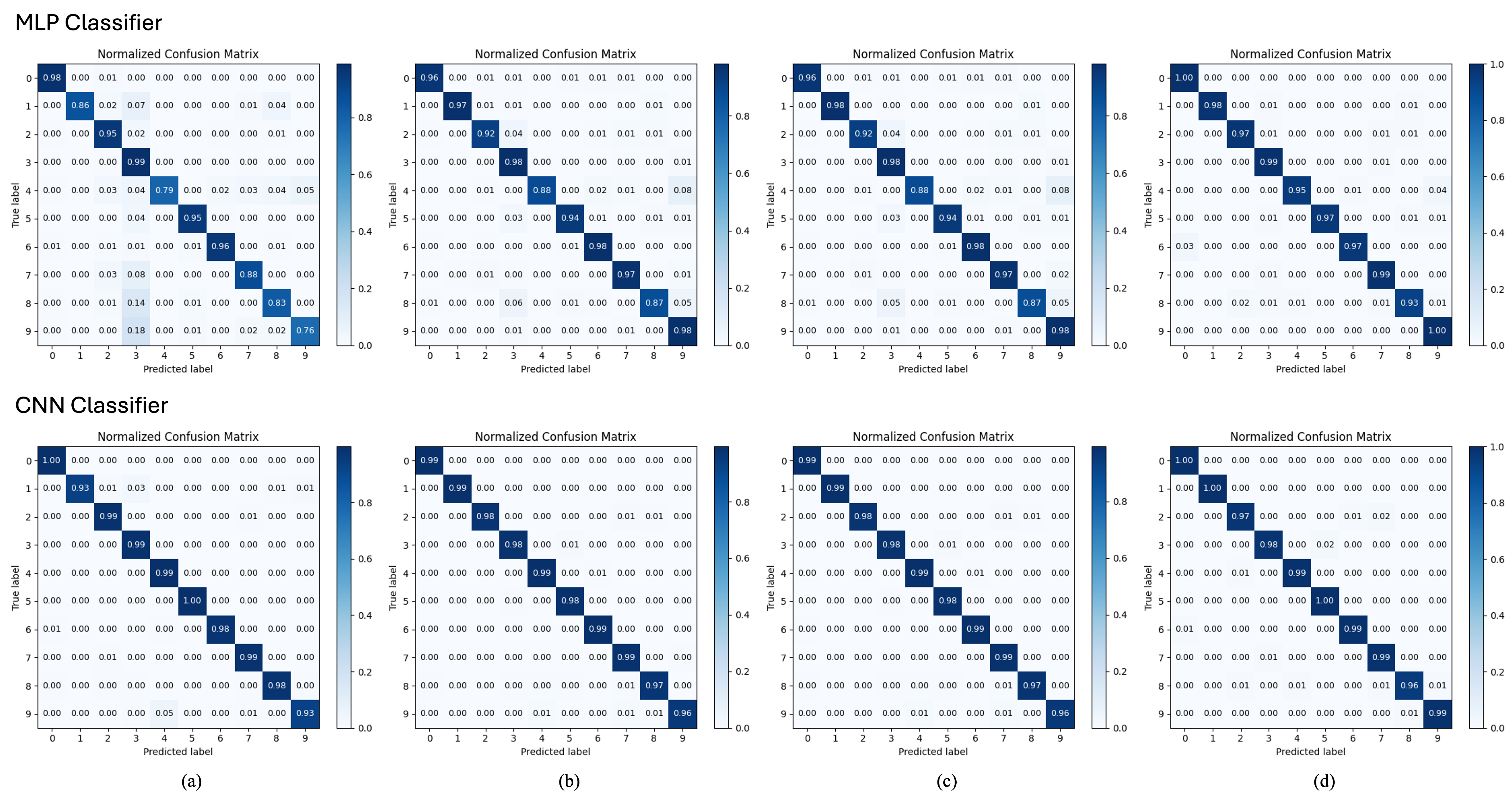}
\caption{Confusion matrices for (a) automatic encoding, (b) TNQE-full, (c) TNQE-core, and (d) TNQE-unitary on the MNIST classification task in \cref{fig:circuit_com}d, using pre-trained MLP (top) and CNN (bottom) classifiers.}
\label{fig:confusion_matrix}
\end{center}
% \vskip -15pt
\end{figure}

\section{More details about experiments in main text}
\label{app:hardware_details}

The detailed confusion matrices in \cref{fig:confusion_matrix} further illustrate the classification performance of different encodings. These findings indicate that the proposed TNQE methods effectively preserve essential image features during encoding, leading to improved classification performance.
Both classifiers are pre-trained on the original MNIST training set without data augmentation and then applied to the approximated images. The CNN consists of Conv($1\!\to\!10$, $5\times5$) $\to$ ReLU $\to$ MaxPool($2\times2$) $\to$ Conv($10\!\to\!20$, $3\times3$) $\to$ ReLU $\to$ FC($2000\!\to\!500$) $\to$ FC($500\!\to\!10$), and the MLP is $784\!\to\!256\!\to\!128\!\to\!10$ with ReLU activations.
As a complementary perceptual measure, \cref{tab_app:lpips} reports the learned perceptual image patch similarity (LPIPS) between the original and approximated images using the standard AlexNet and VGG backbones, together with analogous feature distances computed with our pre-trained classifiers (lower is better). The ordering is identical under all four backbones and consistent with the accuracy and SSIM results: TNQE-unitary is perceptually closest to the original.

\begin{table}[htbp]
  \centering
  \caption{Perceptual distances between original and approximated MNIST images (lower is better).}
  \vskip 9pt
    \begin{tabular}{ccccc}
    \toprule
    Methods & AlexNet & VGG & CNN & MLP \\
    \midrule
    TN reconstruction & 0.090 & 0.150 & 0.531 & 0.050 \\
    Automatic encoding & 0.144 & 0.298 & 0.575 & 0.154 \\
    TNQE-full & 0.098 & 0.183 & 0.551 & 0.071 \\
    TNQE-core & 0.098 & 0.183 & 0.549 & 0.071 \\
    TNQE-unitary & 0.066 & 0.092 & 0.367 & 0.013 \\
    \bottomrule
    \label{tab_app:lpips}
    \end{tabular}
\end{table}

Due to page limitations in the main text, we present additional metrics in \cref{tab:ablation_others}. These results closely match those in \cref{tab:ablation_mse}: increasing the rank generally improves approximation quality but also increases circuit complexity.
Regarding the number of layers in TNQE-unitary, \cref{tab:per_depth} further confirms the observation in \cref{fig:per_depth} that increasing the number of layers generally improves performance. Based on these results, we select rank 4 for TNQE-full and TNQE-core, rank 8 for TNQE-unitary, and four layers for TNQE-unitary in the main experiments.

\begin{table}[htbp]
  \centering
  \caption{Ablation study on different ranks using TNQE on MNIST. The circuit structure is shown in parentheses as (qubits, depth).}
    \vskip 9pt
  \resizebox{\linewidth}{!}{
    \begin{tabular}{cccccccccc}
    \toprule
    \multicolumn{1}{l}{Methods} & Rank  & 1     & 2     & 3     & 4     & 5     & 8     & 9     & 16 \\
    \midrule
    \multirow{4}{*}{Full} & BCE   & 1.137407 & 0.504207 & 0.248129 & 0.125140 & 0.100464 & 0.067908 & 0.065382 & 0.093536 \\
          & PSNR  & 10.9041 & 12.9945 & 15.5608 & 19.0217 & 21.5318 & 29.9344 & 33.4708 & 34.0668 \\
          & SSIM  & 0.287041 & 0.583822 & 0.780099 & 0.906428 & 0.946004 & 0.989724 & 0.993186 & 0.969076 \\
          & Structure & (10, 7) & (11, 195) & (12, 945) & (12, 945) & (13, 4185) & (13, 4185) & (14, 17625) & (14, 17625) \\
    \midrule
    \multirow{4}{*}{Core} & BCE   & 1.150839 & 0.496339 & 0.249040 & 0.124900 & 0.100473 & 0.068111 & 0.065349 & 0.092729 \\
          & PSNR  & 10.8891 & 12.9989 & 15.5508 & 19.0257 & 21.5332 & 29.8655 & 33.2292 & 33.9262 \\
          & SSIM  & 0.283047 & 0.583711 & 0.779720 & 0.906446 & 0.945979 & 0.989542 & 0.993234 & 0.969857 \\
          & Structure & (10, 4) & (18, 24) & (26, 116) & (26, 116) & (31, 242) & (34, 496) & (39, 1006) & (42, 2028) \\
    \midrule
    \multirow{4}{*}{Unitary} & BCE   & 0.583105 & 0.197783 & 0.113675 & 0.113823 & 0.092351 & 0.092333 & 0.083594 & 0.083581 \\
          & PSNR  & 8.2066 & 13.7810 & 18.0264 & 18.0345 & 20.6001 & 20.5837 & 21.9041 & 21.8522 \\
          & SSIM  & 0.184680 & 0.673837 & 0.855882 & 0.855416 & 0.915680 & 0.91590331 & 0.938823 & 0.939511 \\
          & Structure & (10, 8) & (11, 56) & (12, 61) & (12, 61) & (13, 81) & (13, 81) & (14, 82) & (14, 82) \\
    \bottomrule
    \end{tabular}}
  \label{tab:ablation_others}%
\end{table}%

\begin{table}[htbp]
  \centering
  \caption{The impact of number of layers in TNQE-unitary on image approximation quality.}
    \vskip 9pt
    \resizebox{\linewidth}{!}{
    \begin{tabular}{ccccccccccc}
      \toprule
    Layers & 1     & 2     & 3     & 4     & 5     & 6     & 8     & 10     & 14    & 20 \\
      \midrule
    MSE   & 0.045022 & 0.021625 & 0.015167 & 0.011249 & 0.008361 & 0.006421 & 0.003857 & 0.002621 & 0.001147 & 0.000504 \\
    BCE   & 0.202573 & 0.123985 & 0.103954 & 0.092333 & 0.083855 & 0.077791 & 0.069254 & 0.064738 & 0.058689 & 0.055501 \\
    PSNR  & 13.6188 & 17.2647 & 19.1014 & 20.5837 & 21.8803 & 23.1861 & 25.7583 & 27.8214 & 32.8591 & 38.8589 \\
    SSIM  & 0.694205 & 0.840451 & 0.885914 & 0.915903 & 0.938019 & 0.952568 & 0.971858 & 0.981073 & 0.992111 & 0.996577 \\
    \midrule
    Depth & 21    & 41    & 61    & 81    & 101   & 121   & 161   & 201   & 281   & 401 \\
      \bottomrule
    \end{tabular}}
  \label{tab:per_depth}%
\end{table}%

\paragraph{Details of quantum hardware.}
We conduct experiments on IBM Quantum Platform, specifically using four different quantum processing units (QPUs):
ibm\_torino with Heron r1: 133 qubits, 2Q error (median): 2.61E-3, 2Q error (layered): 8.81E-3;
ibm\_fez with Heron r2: 156 qubits, 2Q error (median): 2.57E-3, 2Q error (layered): 4.97E-3;
ibm\_kawasaki with Heron r2: 156 qubits, 2Q error (median): 1.37E-3, 2Q error (layered): 6.60E-3;
ibm\_boston with Heron r3: 156 qubits, 2Q error (median): 1.26E-3, 2Q error (layered): 2.03E-3.
The MNIST evaluation in \cref{tab:real_quantum} uses three of these QPUs, while \texttt{ibm\_kawasaki} is used for the additional resolution experiment in \cref{fig:blob}. Owing to the high cost associated with IBM quantum services, we restrict the MNIST evaluation to the first 100 test samples.
Every circuit is executed with 50,000 shots.
Circuits are transpiled to each device with the Qiskit preset pass manager. The depths reported in the main text are logical depths in a gate set of single-qubit rotations and CNOTs, before this device-specific routing. No error mitigation or hardware-specific optimization is applied.
Nevertheless, a clear performance gap is observed, with our method significantly outperforming both amplitude encoding and automatic encoding, as also illustrated in \cref{fig:real_quantum}.

\paragraph{Implementation details of quantized tensor train optimization.}
Regarding the classical optimization of QTT in TNQE-full and TNQE-core, we employ the Adam optimizer \citep{kingma2014adam} with a learning rate of 0.01, and train for 2000 epochs on MSE loss.
Regarding TNQE-unitary, we adopt more advanced optimization techniques to enhance convergence and stability.
Specifically, we utilize the AdamW optimizer with a maximum learning rate of $0.03$ and a weight decay of $10^{-4}$, combined with a one-cycle learning rate schedule for stable training, and train for 1000 epochs on KL divergence loss.

\begin{figure}[htbp]
\begin{center}
\includegraphics[width=0.7\linewidth]{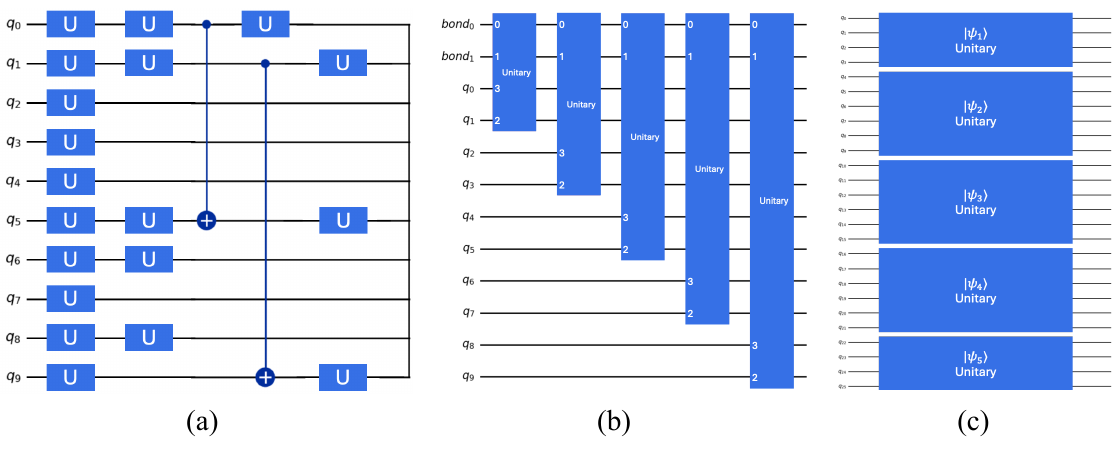}
\caption{Comparison of quantum circuit for (a) amplitude encoding (partially shown due to its depth), (b) TNQE-full encoding and TNQE-unitary encoding, and (c) TNQE-core encoding.}
\vskip -9pt
\label{fig:circuits}
\end{center}
\end{figure}

\paragraph{Differences between encoding circuits.}
For completeness, we provide circuit illustrations corresponding to the encoding strategies discussed in the main text (see Fig.\ref{fig:circuits} in this appendix). These circuits explicitly highlight the structural differences between classical amplitude encoding and the proposed TNQE variants. In particular, Fig.\ref{fig:circuits}a shows the highly entangled global operations required for amplitude encoding, while Fig.~\ref{fig:circuits}b and c illustrate how TNQEs restrict operations to local tensor cores, resulting in more structured and shallow circuit implementations.

\begin{table}[htbp]
  \centering
  \caption{Results with original image as reference.}
    \vskip 9pt
    \begin{tabular}{ccccc}
    \toprule
    methods & MSE   & BCE   & PSNR  & SSIM \\
    \midrule
    Automatic encoding & 0.015656 & 0.147523 & 18.3669 & 0.882500 \\
    TNQE-full & 0.015700 & 0.129583 & 18.6845 & 0.906974 \\
    TNQE-core & 0.015639 & 0.129311 & 18.6903 & 0.907203 \\
    TNQE-unitary & 0.011249 & 0.092333 & 20.5837 & 0.915903 \\
    \bottomrule
    \end{tabular}
    \label{tab_app:org_com}
\end{table}
\begin{table}[htbp]
  \centering
  \caption{Results from each stage.}
    \vskip 9pt
    \begin{tabular}{cccccc}
    \toprule
    \multicolumn{2}{c}{Methods} & \multicolumn{2}{c}{TNQE-full} & \multicolumn{2}{c}{TNQE-core} \\
    \midrule
    Com.  & Img. to TN & TN to QC & Img. To QC & TN to QC & Img. To QC \\
    \midrule
    MSE   & 0.013135 & 0.002234 & 0.015700 & 0.002182 & 0.015639 \\
    BCE   & 0.113215 & 0.165439 & 0.129583 & 0.165311 & 0.129311 \\
    PSNR  & 19.3358 & 28.4850 & 18.6845 & 28.4913 & 18.6903 \\
    SSIM  & 0.924981 & 0.983236 & 0.906974 & 0.983443 & 0.907203 \\
    \bottomrule
    \label{tab_app:stage_com}
    \end{tabular}
\end{table}

\paragraph{Detailed comparison of each stage of TNQEs.}
Based on \cref{tab:quantitative_com}, we further analyze the approximation quality of each stage. Comparison with the original image measures the end-to-end error of the full pipeline, as shown in \cref{tab_app:org_com}. TNQE-full and TNQE-core contain two stages: (i) obtaining a low-rank image representation and (ii) converting its tensor cores into a quantum circuit. Comparing their circuit outputs with the TN-reconstructed images isolates the additional error introduced by the second stage, as shown in \cref{tab_app:stage_com}. The results indicate that most of the error comes from the initial tensor approximation, while circuit construction preserves the TN-reconstructed image.

\section{Additional discussion and experimental results}
\label{app:discussion}

\paragraph{Relation to feature-based quantum-classical hybrid pipelines.}
Several hybrid approaches first use classical models (e.g., CNNs) to extract low-dimensional features and then encode them into a simple quantum circuit for subsequent processing \citep{ranga2024hybrid,lu2025fidelity}.
As discussed in the main text, although such hybrid approaches can in principle handle arbitrary classical data, they rely on pre-trained classical models to extract task-specific feature representations. Recovering the original input typically requires an additional classical decoder, which limits their applicability in truly end-to-end quantum machine learning settings. Moreover, once high-dimensional data are compressed into low-dimensional features, simple classical algorithms often achieve performance that substantially surpasses current quantum methods, thereby weakening the necessity of quantum processing in these pipelines. As a result, these approaches tend to exhibit advantages only for specific tasks and datasets, rather than providing a general framework for quantum machine learning.

TNQE targets a fundamentally different goal. Rather than learning specific task-features, we focus on quantum state preparation from high-dimensional inputs under realistic circuit constraints. Although tensor network decompositions can be interpreted as a form of compression, they differ fundamentally from CNN-style representation learning: QTT relies on deterministic reshaping (quantization) followed by low-rank factorization that captures multi-scale correlations, without being trained to extract high-level semantic features.

Importantly, the quantum circuits produced by TNQE admit approximate reconstruction of the original input through measurements and simple post-processing (e.g., normalization, factorization). This property is generally absent in CNN-based hybrid approaches, where the encoding is not designed to preserve sufficient information for input-level reconstruction.

\label{app:real_quantum}
\begin{figure}[ht]
\begin{center}
\includegraphics[width=0.9\linewidth]{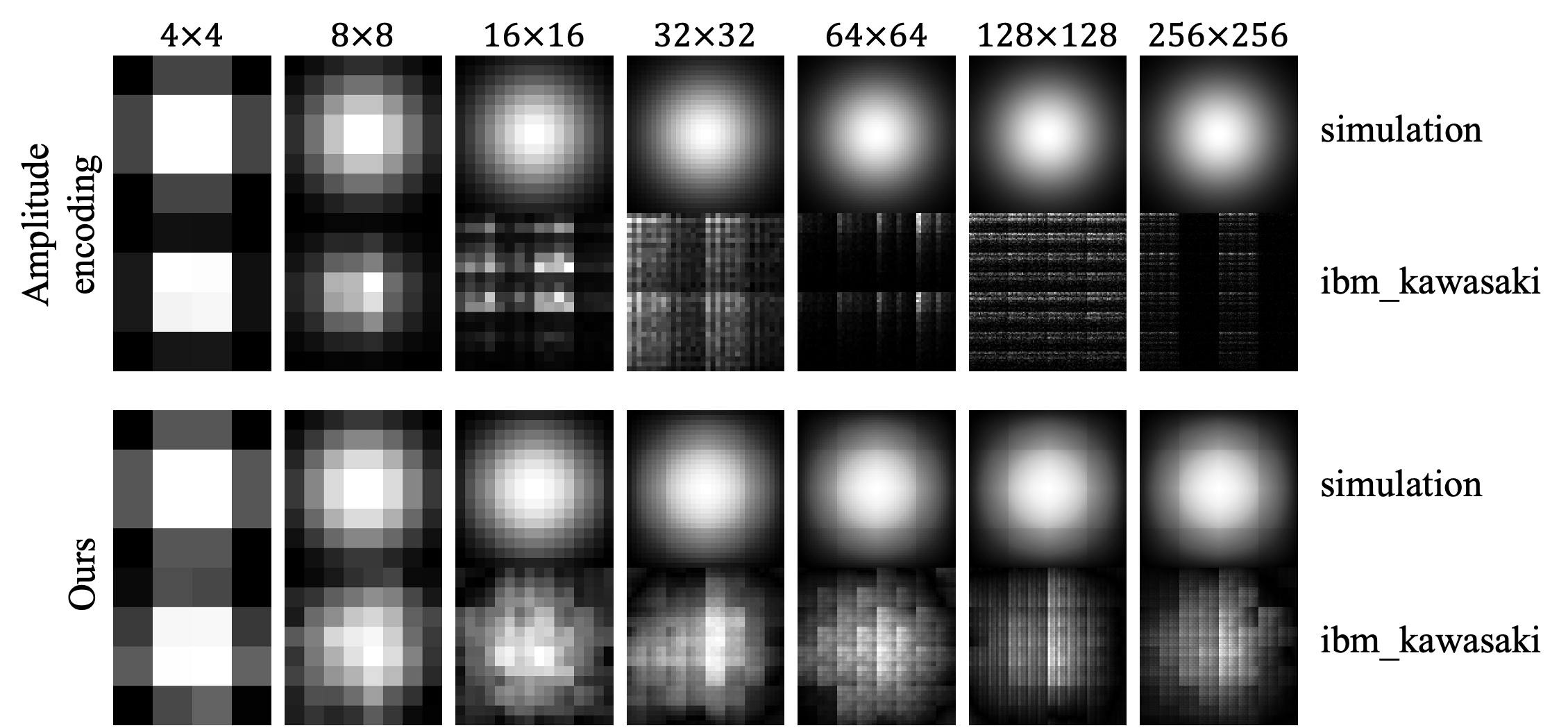}
\caption{Comparison of simulated and real quantum results on ibm\_kawasaki with Heron r2.}
\label{fig:blob}
\end{center}
% \vskip -9pt
\end{figure}

\paragraph{Hardware results and high-resolution stress test.}
We include real quantum experiments to evaluate practical feasibility and to expose how different encoders behave under hardware noise. To this end, we create synthetic images consisting of Gaussian ``blobs'' of varying sizes and encode them using amplitude encoding and TNQE-core with $r=4$. We then execute the resulting circuits on \texttt{ibm\_kawasaki} with Heron r2 and compare the measurement outcomes, as shown in \Cref{fig:blob}. Both amplitude encoding and TNQE reproduce the target in simulation, whereas the hardware outputs degrade substantially as resolution increases. The classical hardware used is an NVIDIA RTX A5000 (24 GB GPU memory) with Qiskit v2.2.3 and PyTorch v2.0.0 in Python v3.10.
Specifically, amplitude encoding performs an exact encoding of values over the entire pixel space and, at the cost of deep circuits, can achieve very high fidelity in noiseless settings. As shown in \Cref{fig:mnist_visual}, its simulated results are almost indistinguishable from the ground truth. TNQE follows a similar objective: in simulation, it enables efficient encoding of high-resolution inputs, and can even represent $256 \times 256$ color images, as demonstrated in \Cref{fig:vis_comparison}.
However, we observe a pronounced gap between simulation and executions on real quantum hardware, which can render some encoding strategies ineffective. This discrepancy is primarily due to hardware noise, which becomes increasingly severe as circuit depth and the number of qubits grow, leading to the accumulation of errors. In practice, amplitude encoding already fails on real hardware for only $16 \times 16$ images.
While TNQE also exhibits noticeable degradation under hardware noise, it retains nontrivial structural information: even at resolutions as high as $256 \times 256$, the approximated results still preserve coarse blob-like contours. These observations suggest that, although current hardware noise remains a major bottleneck, TNQE merits further investigation in future work.

Importantly, in this work we do not explicitly optimize for execution on noisy quantum hardware. Our primary goal is to develop a new quantum data encoding paradigm that enables the construction of structured and circuit-efficient quantum circuits. Nevertheless, despite the absence of any noise-aware training or hardware-specific optimization, the performance of TNQE on real quantum devices already exhibits encouraging potential. This observation suggests that integrating TNQE with noise-aware optimization strategies, error mitigation techniques, and hardware-adaptive compilation in future work could further enhance its performance on real quantum hardware. Such extensions are natural and well aligned with the design of the proposed framework.
Moreover, many existing approaches remain largely at the level of theoretical analysis or simulation. For example, encoding a $256 \times 256$ image with QPIE \citep{brunet2024quantum} requires at least 131,070 quantum operators, while BRQI, GQIR, and NEQR \citep{lisnichenko2023quantum} require circuit depths of 784,866, 392,236, and 392,281, respectively. These resource requirements place such methods far beyond the capabilities of current quantum devices \citep{brunet2024quantum}. Our work takes a concrete step in this direction by explicitly constructing shallower executable circuits.

\begin{table}[htbp]
  \centering
  \caption{Downstream classification accuracy (\%) of pre-trained CNN / MLP classifiers on images decoded from real quantum hardware.}
    \vskip 9pt
    \begin{tabular}{ccccccc}
    \toprule
    \multirow{2}{*}{QPUs} & \multicolumn{3}{c}{CNN} & \multicolumn{3}{c}{MLP} \\
    \cmidrule(lr){2-4} \cmidrule(lr){5-7}
     & Amp. & Auto. & Ours & Amp. & Auto. & Ours \\
    \midrule
    ibm\_torino (Heron r1) & 13 & 81 & 91 & 13 & 43 & 74 \\
    ibm\_fez (Heron r2) & 13 & 78 & 95 & 17 & 43 & 84 \\
    ibm\_boston (Heron r3) & 10 & 87 & 95 & 15 & 76 & 86 \\
    \midrule
    Mean & 12 & 82 & 93.7 & 15 & 54 & 81.3 \\
    \bottomrule
    \label{tab_app:downstream_real}
    \end{tabular}
\end{table}

\paragraph{Downstream task under hardware noise.}
Encoding quality under noise is the precondition for any downstream algorithm, since the downstream task consumes the encoded state as its input: encodings that are essentially exact in simulation (\Cref{fig:mnist_visual}) degrade to noise on hardware (\Cref{tab:real_quantum,fig:real_quantum}), whereas TNQE still preserves the semantic information. To quantify this on a downstream task, we feed the images decoded from the hardware measurements in \Cref{tab:real_quantum} (the first 100 MNIST test images on each QPU) to the CNN and MLP classifiers of \Cref{fig:circuit_com}d, which are pre-trained on the original MNIST data and kept fixed. As shown in \Cref{tab_app:downstream_real}, amplitude encoding drops to chance level on all three QPUs, automatic encoding loses a large part of its accuracy, while TNQE-core retains high accuracy. The encoded states produced by TNQE therefore remain usable for downstream processing under realistic hardware noise.

\paragraph{Additional experiments on CIFAR-10 dataset.}
To further broaden the empirical evaluation, we have conducted additional experiments on CIFAR-10, as shown in \cref{tab_app:cifar}, further supporting that the advantages of TNQE are not limited to MNIST-style data.

\begin{table}[htbp]
  \centering
  \caption{The results on CIFAR-10 dataset.}
    \vskip 9pt
    \begin{tabular}{ccccc}
    \toprule
    Methods & MSE   & BCE   & PSNR  & SSIM \\
    \midrule
    TNQE-full & 0.011282 & 0.532206 & 19.9926 & 0.909408 \\
    TNQE-core & 0.011670 & 0.533020 & 19.9174 & 0.908461 \\
    TNQE-unitary & 0.010712 & 0.530599 & 20.2312 & 0.924487 \\
    \bottomrule
    \label{tab_app:cifar}
    \end{tabular}
    \vskip -9pt
\end{table}
\begin{table}[htbp]
  \centering
  \caption{Computational cost of TNQEs.}
    \vskip 9pt
    \begin{tabular}{cccc}
    \toprule
    Methods & Img. to TN & TN to QC & Inference time \\
    \midrule
    TNQE-full & 0.0118 s & 0.0012 s & 0.5768 s \\
    TNQE-core & 0.0118 s & 0.0005 s & 2.5720 s \\
    TNQE-unitary & 0.2523 s & 0.0008 s & 0.6016 s \\
    \bottomrule
    \label{tab_app:runtime}
    \end{tabular}
\end{table}

\paragraph{Analysis of computational cost.}
Regarding computational cost, we measure the runtime of each stage, including tensor decomposition (Img. to TN), circuit construction (TN to QC), and inference time of measurement, as shown in \cref{tab_app:runtime}.

Importantly, these are one-time costs for preparing the quantum dataset. In this sense, TNQEs follow a data-preparation paradigm similar to preprocessing pipelines, where additional upfront computation is used to obtain a more efficient or structured representation. Regarding potential applications, TNQE-unitary is suitable for preparing quantum datasets, repeated downstream training or evaluation, and studies in which circuit efficiency is the primary bottleneck after encoding.

\paragraph{Comparison with modern quantum circuit encodings.}
As shown in \cref{tab_app:comparison}, we compare the proposed TNQE methods with modern, variational quantum circuit encodings, including flexible representation of quantum images (FRQI), novel enhanced quantum representation (NEQR), quantum probability image encoding (QPIE), and automatic quantum circuit encoding (AQCE). Our methods are more resource-efficient, especially in terms of depth and operators. This is because most of these methods are designed to encode pixel values exactly into the quantum states, which typically leads to complex circuits, whereas TNQE is approximate by design. TNQEs constitute a structured approximate encoding framework that operates on compact tensor representations. By leveraging low-rank tensor decompositions, TNQEs can capture essential features of the input data while significantly reducing the number of parameters and operations required for encoding. This results in shallower circuits that are more amenable to execution on near-term quantum hardware, while still preserving sufficient information for downstream tasks. Moreover, TNQE-unitary explicitly incorporates unitary constraints into the optimization process, ensuring that the resulting circuits are directly realizable without additional post-processing or approximations.

\begin{table}[t]
\centering
\caption{Comparison of different quantum encoding methods.}
  \vskip 9pt
\label{tab_app:comparison}
\begin{tabular}{lccccc}
\toprule
Method & Qubit & Depth & Unitary & CNOT & Complexity \\
\midrule
FRQI \citep{le2011flexible}         & 11 & 283651 & 235530 & 139264 & $O(2^{4n})$ \\
NEQR \citep{zhang2013neqr} & 18 & 12088  & 40334  & 4034   & $O(qn2^{2n})$ \\
QPIE \citep{brunet2024quantum} & 10 & 2028   & 1023   & 1013   & $O(2^n)$ \\
AQCE \citep{shirakawa2024automatic} & 10 & 208    & 371    & 129    & $O(n^3)$ \\
TNQE-full     & 12 & 945    & 900    & 475    & $O(4nr^2)$ \\
TNQE-core     & 26 & 116    & 219    & 193    & $O(4r^2)$ \\
TNQE-unitary  & 13 & 81     & 100    & 120    & $O(nN_l)$ \\
\bottomrule
\end{tabular}
\end{table}

\paragraph{Comparison with MPS/TN-based encoding methods.}
MPS/TN-based encoding methods share a two-stage structure: a classical tensor-network representation is constructed first and then loaded onto qubits. For the representation stage we compare four methods on the first 100 MNIST images: a plain MPS/TT, TT-SVD, PuTT \citep{loeschcke2024coarse} (state-of-the-art, progressive core optimization), and our jointly optimized QTT. As shown in \Cref{tab_app:tn_rep}, all representations reach an MSE of about $10^{-3}$ at $r=8$ and essentially fit MNIST, so representation quality is not the bottleneck of TN-based data loading. What separates methods in quantum data encoding is the conversion stage: the complexity of the resulting circuit, especially its depth, directly determines how much hardware noise accumulates. Loading the $r=8$ representation with amplitude encoding requires a circuit of 10 qubits with depth 2028. \Cref{tab_app:tn_noisy} reports the reconstruction quality of these pipelines when the loading circuits are simulated under two-qubit depolarizing noise at the median error rate of \texttt{ibm\_boston}, together with the same pipelines at smaller ranks whose depths are closest to the depth 81 of TNQE-unitary. Although the classical representations are an order of magnitude more accurate than TNQE-unitary in noiseless reconstruction, this advantage is erased by the deep loading circuit, and even at depths comparable to or larger than 81 these methods still underperform TNQE-unitary.
More fundamentally, the conversion stage dominates the circuit cost regardless of the TN quality: \Cref{tab_app:tn_depth} shows that loading the identical QTT cores through amplitude encoding remains costly across all ranks. This is precisely the motivation for the proposed pipeline. TNQE-unitary avoids the conversion stage entirely: it directly parameterizes each tensor core as a shallow block unitary and optimizes the gate parameters within the predefined circuit architecture (\Cref{subsec:unitary_aware_qtt}), so the learned parameters directly specify executable quantum gates, without post hoc unitary conversion or synthesis.

\begin{table}[htbp]
  \centering
  \caption{Classical reconstruction MSE of different tensor-network representations on the first 100 MNIST images.}
    \vskip 9pt
    \begin{tabular}{ccccc}
    \toprule
    Representation & $r=1$ & $r=2$ & $r=4$ & $r=8$ \\
    \midrule
    MPS/TT & 0.0607 & 0.0317 & 0.0115 & 0.0014 \\
    TT-SVD & 0.0591 & 0.0383 & 0.0093 & 0.0010 \\
    PuTT & 0.0614 & 0.0395 & 0.0092 & 0.0010 \\
    QTT (ours) & 0.0589 & 0.0383 & 0.0092 & 0.0010 \\
    \bottomrule
    \label{tab_app:tn_rep}
    \end{tabular}
    \vskip -9pt
\end{table}
\begin{table}[htbp]
  \centering
  \caption{TN-based encoding pipelines (classical representation + amplitude-encoding loading) versus TNQE-unitary. Depths are logical depths as in \Cref{tab:circuit_comparison}.}
    \vskip 9pt
    \begin{tabular}{cccccccc}
    \toprule
    Methods & Rank & Qubits & Depth & MSE & BCE & PSNR & SSIM \\
    \midrule
    MPS/TT-based & 8 & 10 & 2028 & 0.0149 & 0.1084 & 18.44 & 0.5301 \\
    TT-SVD-based & 8 & 10 & 2028 & 0.0142 & 0.1059 & 18.63 & 0.5513 \\
    PuTT-based & 8 & 10 & 2028 & 0.0143 & 0.1058 & 18.61 & 0.5485 \\
    \midrule
    MPS/TT-based & 1 & 5 & 54 & 0.0606 & 0.5886 & 12.57 & 0.2255 \\
    TT-SVD-based & 1 & 5 & 54 & 0.0591 & 0.5034 & 12.67 & 0.2431 \\
    PuTT-based & 1 & 5 & 54 & 0.0595 & 0.3902 & 12.63 & 0.2555 \\
    \midrule
    MPS/TT-based & 2 & 7 & 242 & 0.0390 & 0.2614 & 14.43 & 0.3359 \\
    TT-SVD-based & 2 & 6 & 116 & 0.0400 & 0.3193 & 14.58 & 0.3986 \\
    PuTT-based & 2 & 6 & 116 & 0.0404 & 0.2886 & 14.32 & 0.3996 \\
    \midrule
    TNQE-unitary & 8 & 13 & 81 & 0.0116 & 0.0934 & 19.91 & 0.6437 \\
    \bottomrule
    \label{tab_app:tn_noisy}
    \end{tabular}
\end{table}

\begin{table}[htbp]
  \centering
  \caption{Circuit structure (qubits, depth) for loading the same QTT cores through amplitude encoding (QTT+AE) versus TNQE-unitary, as a function of the rank.}
    \vskip 9pt
    \begin{tabular}{ccccccc}
    \toprule
    Rank & 1 & 2 & 3 & 4 & 5 & 8 \\
    \midrule
    QTT+AE & (5, 54) & (6, 116) & (8, 496) & (8, 496) & (9, 1006) & (10, 2028) \\
    TNQE-unitary & (10, 8) & (11, 56) & (12, 61) & (12, 61) & (13, 81) & (13, 81) \\
    \bottomrule
    \label{tab_app:tn_depth}
    \end{tabular}
\end{table}

\paragraph{Barren plateau phenomenon or vanishing gradients phenomenon.}
In our current setting, TNQE-unitary is optimized entirely on a classical system: the circuit architecture is fixed in advance, and the block-unitary parameters are trained using standard classical gradient-based optimization. We perform optimization on a classical system, but the optimization objective is not a simulated quantum circuit. Instead, we construct a classical QTT model whose structure follows the same design as the target quantum circuit, and optimize this classical model using gradient descent. Specifically, TNQE-unitary performs a QTT decomposition, $X \approx \mathcal{A}_1 \mathcal{A}_2 \cdots \mathcal{A}_L$, where each tensor core $\mathcal{A}_i$ must satisfy the unitary constraint, rather than being an arbitrary tensor. The specific construction method is presented in Section 4.3. Since each tensor core is composed of unitary operations, the learned parameters can be directly mapped to the corresponding quantum circuit. Moreover, in TNQE-unitary, the optimization is performed on a classical QTT model, not simulated quantum circuit, which does not involve repeated quantum evaluations, shot-based estimation, or direct optimization of a simulated quantum execution pipeline.

Therefore, TNQE-unitary alleviates the conventional barren plateau issue. Regarding the similar vanishing gradients phenomenon in classical optimization, empirically, we do not observe optimization collapse in the regimes studied in this work. TNQE-unitary converged stably across the tested settings, and increasing the rank or number of layers consistently improved reconstruction quality.

\begin{table}[htbp]
  \centering
  \caption{Results over five runs (mean and standard deviation).}
    \vskip 9pt
  \resizebox{\linewidth}{!}{
    \begin{tabular}{cccccccc}
    \toprule
    Methods & No.   & 1     & 2     & 3     & 4     & 5     & Mean $\pm$ Std.  \\
    \midrule
    \multirow{4}[0]{*}{TNQE-full} & MSE   & 0.014735 & 0.014199 & 0.013841 & 0.013850 & 0.014348 & 0.014194 $\pm$ 0.000374 \\
          & BCE   & 0.125140 & 0.122135 & 0.121945 & 0.122370 & 0.121986 & 0.122715 $\pm$ 0.001366 \\
          & PSNR  & 19.0217 & 19.1892 & 19.3041 & 19.4389 & 19.2148 & 19.2337 $\pm$ 0.1536 \\
          & SSIM  & 0.906428 & 0.908388 & 0.910602 & 0.911823 & 0.905230 & 0.908494 $\pm$ 0.002760 \\
          \midrule
    \multirow{4}[0]{*}{TNQE-core} & MSE   & 0.014731 & 0.014174 & 0.013878 & 0.013858 & 0.014342 & 0.014197 $\pm$ 0.000362 \\
          & BCE   & 0.124900 & 0.122060 & 0.124593 & 0.121655 & 0.122007 & 0.123043 $\pm$ 0.001566 \\
          & PSNR  & 19.0257 & 19.2072 & 19.2668 & 19.4386 & 19.2243 & 19.2325 $\pm$ 0.1476 \\
          & SSIM  & 0.906446 & 0.908707 & 0.910152 & 0.911616 & 0.905427 & 0.908469 $\pm$ 0.002557 \\
    \midrule
    \multirow{4}[1]{*}{TNQE-unitary} & MSE   & 0.011249 & 0.010985 & 0.010906 & 0.010593 & 0.010318 & 0.010810 $\pm$ 0.000361 \\
          & BCE   & 0.092333 & 0.090795 & 0.090250 & 0.089489 & 0.088213 & 0.090216 $\pm$ 0.001530 \\
          & PSNR  & 20.5837 & 20.6658 & 20.7993 & 20.7021 & 21.0665 & 20.7635 $\pm$ 0.1862 \\
          & SSIM  & 0.915903 & 0.915154 & 0.915879 & 0.920554 & 0.920404 & 0.917579 $\pm$ 0.002665 \\
    \bottomrule
    \end{tabular}}
  \label{tab_app:five_runs}
\end{table}

\paragraph{Statistical stability.}
To further validate the stability of our method, we repeat the experiments five additional times on different subsets with 100 samples, as shown in \cref{tab_app:five_runs}. The results demonstrate that our method consistently achieves robust performance with minimal variance across different runs, underscoring its reliability and effectiveness in quantum data encoding.

\paragraph{Comparison with PCA.}
We perform principal component analysis (PCA) on the original image, retain the top one to five principal components, and then apply amplitude encoding as a low-rank compression baseline. As shown in \cref{tab_app:pca_com}, PCA + AE can also achieve low-rank compression, but its performance is significantly worse than our methods.

\renewcommand{\arraystretch}{0.8}
\begin{table}[htbp]
  \centering
  \caption{Comparison with PCA + amplitude encoding pipeline.}
    \vskip 9pt
    \begin{tabular}{ccccccc}
    \toprule
    Methods & Rank / Com.  & 1     & 2     & 3     & 4     & 5 \\
    \midrule
    \multirow{5}[2]{*}{TNQE-full} & MSE   & 0.088905 & 0.056156 & 0.032136 & 0.014735 & 0.008940 \\
          & BCE   & 1.137407 & 0.504207 & 0.248129 & 0.125140 & 0.100464 \\
          & PSNR  & 10.9041 & 12.9945 & 15.5608 & 19.0217 & 21.5318 \\
          & SSIM  & 0.287041 & 0.583822 & 0.780099 & 0.906428 & 0.946004 \\
          & Circuit & (10, 7) & (11, 195) & (12, 945) & (12, 945) & (13, 4185) \\
    \midrule
    \multirow{5}[2]{*}{TNQE-core} & MSE   & 0.089138 & 0.056174 & 0.032164 & 0.014731 & 0.008943 \\
          & BCE   & 1.150839 & 0.496339 & 0.249040 & 0.124900 & 0.100473 \\
          & PSNR  & 10.8891 & 12.9989 & 15.5508 & 19.0257 & 21.5332 \\
          & SSIM  & 0.283047 & 0.583711 & 0.779720 & 0.906446 & 0.945979 \\
          & Circuit & (10, 4) & (18, 24) & (26, 116) & (26, 116) & (31, 242) \\
    \midrule
    \multirow{5}[2]{*}{TNQE-unitary} & MSE   & 0.162642 & 0.045030 & 0.018398 & 0.018493 & 0.011287 \\
          & BCE   & 0.583105 & 0.197783 & 0.113675 & 0.113823 & 0.092351 \\
          & PSNR  & 8.2066 & 13.7810 & 18.0264 & 18.0345 & 20.6001 \\
          & SSIM  & 0.184680 & 0.673837 & 0.855882 & 0.855416 & 0.915680 \\
          & Circuit & (10, 8) & (11, 56) & (12, 61) & (12, 61) & (13, 81) \\
    \midrule
    \multirow{5}[2]{*}{PCA+AE} & MSE   & 0.052227 & 0.058277 & 0.046965 & 0.039106 & 0.028063 \\
          & BCE   & 0.247707 & 0.292001 & 0.266299 & 0.243545 & 0.208624 \\
          & PSNR  & 13.3882 & 12.9461 & 13.9183 & 15.9557 & 17.7838 \\
          & SSIM  & 0.609144 & 0.578496 & 0.628756 & 0.677571 & 0.739779 \\
          & Circuit & (6, 116) & (7, 242) & (8, 496) & (8, 496) & (9, 1006) \\
    \bottomrule
    \label{tab_app:pca_com}
    \end{tabular}
\end{table}

In contrast, TNQE-unitary achieves significantly lower circuit depth than PCA + AE baseline, while using a comparable number of qubits, and achieves much better quantum data encoding performance. This result further supports the advantages of the proposed TNQE framework.

\paragraph{Dependence on compressibility.}
The advantage of TNQE depends on how well the input is captured by a low-rank QTT, which defines the applicability boundary of the method. \Cref{tab_app:pca_com} already shows that compression alone does not produce this advantage: at the same compression level, PCA + amplitude encoding requires depth 496 with MSE 0.0391, whereas TNQE-unitary with $r=4$ requires depth 61 with MSE 0.0185. To characterize the boundary further, we conduct a stress test spanning MNIST, grayscale CIFAR-10, synthetic high-frequency textures (checkerboards and two-dimensional sinusoids), and uniform random noise, and report in \Cref{tab_app:compressibility} the MSE as a function of the rank for the classical TT approximation and for TNQE-unitary ($N_\ell = 4$).
Two conclusions follow. First, together with \Cref{tab_app:pca_com}, the advantage of TNQE is not an artifact of compression: the structured unitary construction achieves a substantially better fidelity-depth operating point than generic low-rank compression. Second, the dependence on compressibility is shared by any low-rank representation: for uniform noise the classical TT approximation itself stalls at an MSE of 0.0637 at $r=4$, and TNQE-unitary obtains the same error. In other words, TNQE inherits the limitation of the underlying low-rank representation rather than introducing additional approximation error during circuit construction.

\begin{table}[htbp]
  \centering
  \caption{MSE as a function of the rank $r$ for the classical TT approximation and for TNQE-unitary.}
    \vskip 9pt
    \begin{tabular}{ccccccccc}
    \toprule
    \multirow{2}{*}{$r$} & \multicolumn{4}{c}{TT} & \multicolumn{4}{c}{TNQE-unitary} \\
    \cmidrule(lr){2-5} \cmidrule(lr){6-9}
     & MNIST & CIFAR-10 & Texture & Noise & MNIST & CIFAR-10 & Texture & Noise \\
    \midrule
    1 & 0.0598 & 0.0279 & 0.1651 & 0.0821 & 0.1626 & 0.0272 & 0.1569 & 0.0820 \\
    2 & 0.0385 & 0.0160 & 0.0416 & 0.0782 & 0.0450 & 0.0118 & 0.0053 & 0.0734 \\
    4 & 0.0100 & 0.0054 & 0.0000 & 0.0637 & 0.0185 & 0.0068 & 0.0001 & 0.0637 \\
    8 & 0.0010 & 0.0015 & 0.0000 & 0.0386 & 0.0112 & 0.0048 & 0.0001 & 0.0547 \\
    \bottomrule
    \label{tab_app:compressibility}
    \end{tabular}
\end{table}

\paragraph{Trade-offs between fidelity, resources, and cost.}
Each component of the trade-off is quantified separately in the paper: approximation error versus rank (\Cref{tab:ablation_mse,tab:ablation_others}), depth and resource scaling (\Cref{tab:circuit_comparison}, \Cref{fig:circuit_com}a-c, \Cref{tab:per_depth,tab_app:comparison}), training overhead (\Cref{tab_app:runtime}), and downstream classification (\Cref{fig:circuit_com}d, \Cref{fig:confusion_matrix}). \Cref{tab_app:tradeoff} reports them jointly for the main configurations on MNIST.

\begin{table}[htbp]
  \centering
  \small
  \caption{Joint view of the trade-offs on MNIST.}
    \vskip 9pt
    \begin{tabular}{ccccccccc}
    \toprule
    Methods & Qubits & Depth & Ops. & MSE & SSIM & Img. to TN & CNN (\%) & MLP (\%) \\
    \midrule
    TNQE-full & 12 & 945 & 1375 & 0.0157 & 0.9070 & 0.0118 s & 98.22 & 94.48 \\
    TNQE-core & 26 & 116 & 412 & 0.0156 & 0.9072 & 0.0118 s & 98.23 & 94.53 \\
    TNQE-unitary & 13 & 81 & 220 & 0.0112 & 0.9159 & 0.2523 s & 98.70 & 97.60 \\
    \bottomrule
    \label{tab_app:tradeoff}
    \end{tabular}
\end{table}

\paragraph{Limitations and future directions.}
While TNQE demonstrates strong empirical performance and practical scalability, several open questions remain.
First, the unitary-aware optimization in TNQE-unitary is primarily designed to ensure circuit realizability and explicit resource control. A more complete theoretical characterization of its optimization landscape, convergence behavior, and stability would further strengthen the understanding of when and why such parameterizations are effective, and is left to future work.
Second, continued progress in tensor decomposition algorithms, adaptive rank selection, and low-rank representations is therefore expected to translate directly into improved encoding fidelity within the TNQE framework.

\paragraph{Impact statement.}
This paper aims to advance quantum machine learning (QML) by developing more circuit-efficient quantum data encoding methods. By leveraging tensor network representations, TNQE reduces circuit depth and provides explicit control over qubit and gate resources, which contributes to more scalable and hardware-compatible QML pipelines. Many QML algorithms perform well in noiseless simulation but degrade on real quantum hardware due to noise and circuit-depth constraints. TNQE can help mitigate these issues and enable more effective use of near-term quantum devices, which has positive implications for the practical advancement of QML.

%%%%%%%%%%%%%%%%%%%%%%%%%%%%%%%%%%%%%%%%%%%%%%%%%%%%%%%%%%%%

\newpage
\section*{NeurIPS Paper Checklist}

\begin{enumerate}

\item {\bf Claims}
    \item[] Question: Do the main claims made in the abstract and introduction accurately reflect the paper's contributions and scope?
    \item[] Answer: \answerYes{} % Replace by \answerYes{}, \answerNo{}, or \answerNA{}.
    \item[] Justification: The paper has accurately stated the generalization challenges in quantum data encoding tasks and the corresponding technical issues in the abstract and introduction.

\item {\bf Limitations}
    \item[] Question: Does the paper discuss the limitations of the work performed by the authors?
    \item[] Answer: \answerYes{} % Replace by \answerYes{}, \answerNo{}, or \answerNA{}.
    \item[] Justification: The paper has included a "Limitations" section and provided further related discussions in the Appendix.

\item {\bf Theory assumptions and proofs}
    \item[] Question: For each theoretical result, does the paper provide the full set of assumptions and a complete (and correct) proof?
    \item[] Answer: \answerNA{} % Replace by \answerYes{}, \answerNo{}, or \answerNA{}.
    \item[] Justification: The paper does not include theoretical results.

    \item {\bf Experimental result reproducibility}
    \item[] Question: Does the paper fully disclose all the information needed to reproduce the main experimental results of the paper to the extent that it affects the main claims and/or conclusions of the paper (regardless of whether the code and data are provided or not)?
    \item[] Answer: \answerYes{} % Replace by \answerYes{}, \answerNo{}, or \answerNA{}.
    \item[] Justification: The paper fully disclose all the information needed to reproduce the main experimental results of the paper.

\item {\bf Open access to data and code}
    \item[] Question: Does the paper provide open access to the data and code, with sufficient instructions to faithfully reproduce the main experimental results, as described in supplemental material?
    \item[] Answer: \answerYes{} % Replace by \answerYes{}, \answerNo{}, or \answerNA{}.
    \item[] Justification: The code is available at https://github.com/glin2022/tnqe

\item {\bf Experimental setting/details}
    \item[] Question: Does the paper specify all the training and test details (e.g., data splits, hyperparameters, how they were chosen, type of optimizer) necessary to understand the results?
    \item[] Answer: \answerYes{} % Replace by \answerYes{}, \answerNo{}, or \answerNA{}.
    \item[] Justification: The paper specifies all the experiment details.

\item {\bf Experiment statistical significance}
    \item[] Question: Does the paper report error bars suitably and correctly defined or other appropriate information about the statistical significance of the experiments?
    \item[] Answer: \answerYes{} % Replace by \answerYes{}, \answerNo{}, or \answerNA{}.
    \item[] Justification: We report performance across different test samples.

\item {\bf Experiments compute resources}
    \item[] Question: For each experiment, does the paper provide sufficient information on the computer resources (type of compute workers, memory, time of execution) needed to reproduce the experiments?
    \item[] Answer: \answerYes{} % Replace by \answerYes{}, \answerNo{}, or \answerNA{}.
    \item[] Justification: We have provided the information on the computer resources needed (including classical and quantum resources) in Appendix E.

\item {\bf Code of ethics}
    \item[] Question: Does the research conducted in the paper conform, in every respect, with the NeurIPS Code of Ethics \url{https://neurips.cc/public/EthicsGuidelines}?
    \item[] Answer: \answerYes{} % Replace by \answerYes{}, \answerNo{}, or \answerNA{}.
    \item[] Justification: The research conducted in the paper conforms with the NeurIPS Code of Ethics.

\item {\bf Broader impacts}
    \item[] Question: Does the paper discuss both potential positive societal impacts and negative societal impacts of the work performed?
    \item[] Answer: \answerYes{} % Replace by \answerYes{}, \answerNo{}, or \answerNA{}.
    \item[] Justification: We aim to improve advanced quantum data encoding technique, which has positive societal impacts.

\item {\bf Safeguards}
    \item[] Question: Does the paper describe safeguards that have been put in place for responsible release of data or models that have a high risk for misuse (e.g., pre-trained language models, image generators, or scraped datasets)?
    \item[] Answer: \answerNA{} % Replace by \answerYes{}, \answerNo{}, or \answerNA{}.
    \item[] Justification: The paper poses no such risks.

\item {\bf Licenses for existing assets}
    \item[] Question: Are the creators or original owners of assets (e.g., code, data, models), used in the paper, properly credited and are the license and terms of use explicitly mentioned and properly respected?
    \item[] Answer: \answerYes{} % Replace by \answerYes{}, \answerNo{}, or \answerNA{}.
    \item[] Justification: We have cited and properly respected the existing assets in the paper.

\item {\bf New assets}
    \item[] Question: Are new assets introduced in the paper well documented and is the documentation provided alongside the assets?
    \item[] Answer: \answerNA{} % Replace by \answerYes{}, \answerNo{}, or \answerNA{}.
    \item[] Justification: The paper does not release new assets.

\item {\bf Crowdsourcing and research with human subjects}
    \item[] Question: For crowdsourcing experiments and research with human subjects, does the paper include the full text of instructions given to participants and screenshots, if applicable, as well as details about compensation (if any)?
    \item[] Answer: \answerNA{} % Replace by \answerYes{}, \answerNo{}, or \answerNA{}.
    \item[] Justification: The paper does not involve crowdsourcing nor research with human subjects.

\item {\bf Institutional review board (IRB) approvals or equivalent for research with human subjects}
    \item[] Question: Does the paper describe potential risks incurred by study participants, whether such risks were disclosed to the subjects, and whether Institutional Review Board (IRB) approvals (or an equivalent approval/review based on the requirements of your country or institution) were obtained?
    \item[] Answer: \answerNA{} % Replace by \answerYes{}, \answerNo{}, or \answerNA{}.
    \item[] Justification: The paper does not involve crowdsourcing nor research with human subjects.

\item {\bf Declaration of LLM usage}
    \item[] Question: Does the paper describe the usage of LLMs if it is an important, original, or non-standard component of the core methods in this research? Note that if the LLM is used only for writing, editing, or formatting purposes and does \emph{not} impact the core methodology, scientific rigor, or originality of the research, declaration is not required.
    %this research?
    \item[] Answer: \answerNA{} % Replace by \answerYes{}, \answerNo{}, or \answerNA{}.
    \item[] Justification: The core method development in this research does not involve LLMs as any important, original, or non-standard components.

\end{enumerate}

\end{document}